\documentstyle[12pt]{article}

\def\hybrid{\topmargin -20pt	\oddsidemargin 0pt
	\headheight 0pt	\headsep 0pt
	\textwidth 6.25in	
	\textheight 9.5in	
	\marginparwidth .875in
	\parskip 5pt plus 1pt	\jot = 1.5ex}

\hybrid

\def\baselinestretch{1.2}

\catcode`\@=11

\def\marginnote#1{}
\newcount\hour
\newcount\minute
\newtoks\amorpm
\hour=\time\divide\hour by60
\minute=\time{\multiply\hour by60 \global\advance\minute by-\hour}
\edef\standardtime{{\ifnum\hour<12 \global\amorpm={am}%
	\else\global\amorpm={pm}\advance\hour by-12 \fi
	\ifnum\hour=0 \hour=12 \fi
	\number\hour:\ifnum\minute<10 0\fi\number\minute\the\amorpm}}
\edef\militarytime{\number\hour:\ifnum\minute<10 0\fi\number\minute}

\def\draftlabel#1{{\@bsphack\if@filesw {\let\thepage\relax
   \xdef\@gtempa{\write\@auxout{\string
      \newlabel{#1}{{\@currentlabel}{\thepage}}}}}\@gtempa
   \if@nobreak \ifvmode\nobreak\fi\fi\fi\@esphack}
	\gdef\@eqnlabel{#1}}
\def\@eqnlabel{}
\def\@vacuum{}
\def\draftmarginnote#1{\marginpar{\raggedright\scriptsize\tt#1}}

\def\draft{\oddsidemargin -.5truein
	\def\@oddfoot{\sl preliminary draft \hfil
	\rm\thepage\hfil\sl\today\quad\militarytime}
	\let\@evenfoot\@oddfoot	\overfullrule 3pt
	\let\label=\draftlabel
	\let\marginnote=\draftmarginnote
   \def\@eqnnum{(\theequation)\rlap{\kern\marginparsep\tt\@eqnlabel}%
\global\let\@eqnlabel\@vacuum}  }

\def\preprint{\twocolumn\sloppy\flushbottom\parindent 2em
	\leftmargini 2em\leftmarginv .5em\leftmarginvi .5em
	\oddsidemargin -.5in	\evensidemargin -.5in
	\columnsep .4in	\footheight 0pt
	\textwidth 10.in	\topmargin  -.4in
	\headheight 12pt \topskip .4in
	\textheight 6.9in \footskip 0pt
	\def\@oddhead{\thepage\hfil\addtocounter{page}{1}\thepage}
	\let\@evenhead\@oddhead	\def\@oddfoot{}	\def\@evenfoot{} }

\def\numberbysection{\@addtoreset{equation}{section}
	\def\theequation{\thesection.\arabic{equation}}}

\def\underline#1{\relax\ifmmode\@@underline#1\else
	$\@@underline{\hbox{#1}}$\relax\fi}

\def\titlepage{\@restonecolfalse\if@twocolumn\@restonecoltrue\onecolumn
     \else \newpage \fi \thispagestyle{empty}\c@page\z@
	\def\thefootnote{\fnsymbol{footnote}} }

\def\endtitlepage{\if@restonecol\twocolumn \else \newpage \fi
	\def\thefootnote{\arabic{footnote}}
	\setcounter{footnote}{0}}  

\catcode`@=12
\relax

\def\figcap{\section*{Figure Captions\markboth
	{FIGURECAPTIONS}{FIGURECAPTIONS}}\list
	{Figure \arabic{enumi}:\hfill}{\settowidth\labelwidth{Figure 999:}
	\leftmargin\labelwidth
	\advance\leftmargin\labelsep\usecounter{enumi}}}
 \relax
\def\tablecap{\section*{Table Captions\markboth
	{TABLECAPTIONS}{TABLECAPTIONS}}\list
	{Table \arabic{enumi}:\hfill}{\settowidth\labelwidth{Table 999:}
	\leftmargin\labelwidth
	\advance\leftmargin\labelsep\usecounter{enumi}}}
 \relax
\def\reflist{\section*{References\markboth
	{REFLIST}{REFLIST}}\list
	{[\arabic{enumi}]\hfill}{\settowidth\labelwidth{[999]}
	\leftmargin\labelwidth
	\advance\leftmargin\labelsep\usecounter{enumi}}}
 \relax
\makeatletter
\newcounter{pubctr}
\def\publist{\@ifnextchar[{\@publist}{\@@publist}}
\def\@publist[#1]{\list
	{[\arabic{pubctr}]\hfill}{\settowidth\labelwidth{[999]}
	\leftmargin\labelwidth
	\advance\leftmargin\labelsep
	\@nmbrlisttrue\def\@listctr{pubctr}
	\setcounter{pubctr}{#1}\addtocounter{pubctr}{-1}}}
\def\@@publist{\list
	{[\arabic{pubctr}]\hfill}{\settowidth\labelwidth{[999]}
	\leftmargin\labelwidth
	\advance\leftmargin\labelsep
	\@nmbrlisttrue\def\@listctr{pubctr}}}
 \relax
\makeatother
\newskip\humongous \humongous=0pt plus 1000pt minus 1000pt

\newif\ifdtup

\relax

\def\s{\sigma}
\def\thefootnote{\fnsymbol{footnote}}
\def\be{\begin{equation}}
\def\ee{\end{equation}}
\def\ba{\begin{eqnarray}}
\def\ea{\end{eqnarray}}

\def\S{\Sigma}

\def\a{\alpha}

\begin{document}
\renewcommand{\theequation}{\thesection.\arabic{equation}}
\newcommand{\beq}{\begin{equation}}
\newcommand{\eeq}[1]{\label{#1}\end{equation}}
\newcommand{\ber}{\begin{eqnarray}}
\newcommand{\eer}[1]{\label{#1}\end{eqnarray}}
\begin{titlepage}
\begin{center}

\hfill CERN--TH.7047/93\\
\hfill hep-th/9310122\\

\vskip .4in

{\large \bf CONSERVATION LAWS AND GEOMETRY OF PERTURBED
COSET MODELS}

\vskip .7in

{\bf Ioannis Bakas}
\footnote{Permanent address: Department of Physics, University of
Crete,
GR--71409 Heraklion, Greece}
\footnote{e-mail address: BAKAS@SURYA11.CERN.CH}\\
\vskip .1in

{\em Theory Division\\
     CERN\\
     CH--1211 Geneva 23\\
     SWITZERLAND}\\

\vskip .1in

\end{center}

\vskip .7in

\begin{center} {\bf ABSTRACT } \end{center}
\begin{quotation}\noindent
We present a Lagrangian description of the $SU(2)/U(1)$ coset model
perturbed by its first thermal operator. This is the simplest
perturbation
that changes sign under Krammers--Wannier duality.
The resulting theory, which
is a 2--component generalization of the sine--Gordon model, is then
taken in Minkowski space. For negative values of the coupling
constant $g$, it is classically equivalent to the $O(4)$ non--linear
$\s$--model reduced in a certain frame. For $g > 0$,
it describes the relativistic motion of vortices in a constant
external
field. Viewing the classical equations of motion as a zero curvature
condition, we obtain recursive relations for the infinitely many
conservation laws
by the abelianization method of gauge connections.
The higher spin currents are constructed entirely using an
off--critical generalization of the $W_{\infty}$ generators.
We give a geometric interpretation to the corresponding charges in
terms of embeddings. Applications to the
chirally invariant $U(2)$ Gross--Neveu model are
also discussed.
\end{quotation}
\vskip .4cm
CERN--TH.7047/93 \\
October 1993 \\

\end{titlepage}
\vfill
\eject
\def\baselinestretch{1.2}
\baselineskip 16 pt
\setcounter{section}{0}
\section{\bf Introduction}
\noindent
Integrable perturbations of 2--dim conformal field theories (CFT)
have
been studied extensively over the last few years. Zamolodchikov
proved
that there are perturbations of the form
\be
S = S_{CFT} + g ~ \int \Phi
\ee
which take us away from the critical point, but the resulting theory
still
posesses an infinite number of local conservation laws [1, 2]. $\Phi$
is
typically a local spinless field with conformal dimension
${\Delta}_{\Phi}$ which is
found in the operator algebra of the unperturbed theory
$S_{CFT}$. The coupling constant $g$ is a dimensionful parameter of
weight
$(1-\Delta_{\Phi} , 1-\Delta_{\Phi})$.
In these cases, away from criticality, the chiral conservation
laws of the CFT are replaced by
\be
\bar{\partial} A_{s} + \partial B_{s} = 0
\ee
for appropriately chosen non--chiral currents $\{A_{s}\}$ and
$\{B_{s}\}$.
Their existence is deeply related to the null vector conditions on
the
primary field $\Phi$ driving the perturbation. Then, the integrals of
motion off--criticality are
\be
Q_{s} = \int dz A_{s} - d \bar{z} B_{s},
\ee
with $s$ ranging over an infinite set of values, depending on the
model.

In this paper we study the geometry and the classical
conservation laws of the simplest
$Z_{N}$--invariant CFT, ie the $SU(2)/U(1)$ parafermion coset model,
perturbed by its first thermal operator $\Phi = {\epsilon}_{1}$. In
the
large $N$ limit, where all our work will be concentrated, this theory
is
described by the action
\be
S = \int {\partial u \bar{\partial} \bar{u} + \partial \bar{u}
\bar{\partial} u \over 1 - {\mid u \mid}^{2}} + g ~ (1 - {\mid u
\mid}^{2}),
\ee
with ${\mid u \mid}^{2} \leq 1$. The first term is the usual
classical
action for the $SU(2)/U(1)$ coset model and can be obtained from the
$SU(2)$ gauged WZW model after performing the neccessary gauge field
integrations [3, 4, 5]. The potential term, which takes us away from
the
critical point while preserving the $U(1)$ invariance of the theory,
has
conformal dimension 0 ($ = lim_{N \rightarrow \infty} (2 / N + 2)$)
and it corresponds to the
first thermal operator of the unperturbed
model. Many aspects of this model have been
studied before, in connection with the
relativistic theory of vortices in superfluids and other problems
in field theory.

The classical equations of motion that follow from the action (1.4)
are
\be
\partial \bar{\partial} u + {\bar{u} \partial u \bar{\partial} u
\over
1 - {\mid u \mid}^{2}} + g ~ u (1 - {\mid u \mid}^{2}) = 0,
\ee
\be
\partial \bar{\partial} \bar{u} + {u \partial \bar{u} \bar{\partial}
\bar{u} \over 1 - {\mid u \mid}^{2}} + g ~ \bar{u} (1 - {\mid u
\mid}^{2})
= 0.
\ee
This theory provides a 2--component generalization of the
sine--Gordon
model. Indeed, for $u = \bar{u} = cos \theta$, the classical
equations
of motion reduce to
\be
2 ~ \partial \bar{\partial} \theta = g ~ sin 2 \theta
\ee
and so the integrability of the full theory should generalize the
results
we already know for the sine--Gordon model.

Integrable deformations of $Z_{N}$--symmetric models of CFT have been
studied extensively [6, 7]. However, the Lagrangian description and
the
geometric interpretation of these perturbations have not been
addressed
in all generality.
We focus on the simplest perturbation that changes
sign under Krammers--Wannier duality and examine its geometry,
in the context of perturbed
parafermion models. The recent developments in the geometric
interpretation of various CFT coset models as exact theories of
black holes [5], provide the main motivation for considering this
question with a perturbation switched on.
We also write the classical equations of
motion as a zero curvature condition with spectral parameter. Then,
the
abelianization method of gauge connections
[8, 9] is employed to construct classically
the infinitely many local conservation laws of the theory
away from criticality, in a systematic way.
Generalizations to other perturbed coset models
are possible, but go beyond the scope of the present work.

The Lagrangian approach to perturbed CFT coset models has been proven
useful in other occassions for studying the integrability
aspects of some special operators.
For example, when a relevant perturbation by the (1, 3) operator is
applied to the minimal models of the Virasoro algebra with $c \leq
1$,
the perturbed system is effectively described by the sine--Gordon
equation
(1.7), for appropriately chosen values of the coupling constant $g$
[10].
In this case, the local conservation laws can be constructed
systematically
from the zero curvature formulation of the problem and they are
related,
as it is well known, with the Hamiltonian densities of the KdV
hierarchy
(ie, their flows are mutually commuting). More generally, affine Toda
theories
have provided a Lagrangian framework for looking at the problem of
various integrable perturbations (see for instance [11]).

The perturbed theory (1.4) has an interesting physical and
geometric interpretation, when it is defined on Minkowski space.
Based on some old work by Lund and Regge [12,
13], we find that for $g < 0$ its classical equations of motion
describe
a reduced form of the $O(4)$ non--linear $\s$--model
in two dimensions (see also [14]).
For $g > 0$, it describes in a certain gauge the relativistic
motion of vortices in constant external field. To put it differently,
the
physical picture for $g > 0$ is that of a 4--dim bosonic string
propagating in an
axionic background of vortex type. At the conformal point
$g = 0$, it describes the (transverse modes of the) free
Nambu--Goto string in 4--dim Minkowski space in the orthonormal gauge
(see also [15]).

We should mention for completeness (and as independent motivation)
that the
perturbed parafermion theory (1.4) is also interesting for the
classical
problem of two massless Fermi fields
with contact 4--fermion interaction
$(\bar{\psi}^{\a} \psi^{\a})^{2} - (\bar{\psi}^{\a} \gamma_{5}
\psi^{\a})^{2}$ [16].
The Nambu--Jona-Lasinio model in 2--dim Minkowski space
(or Gross--Neveu model) provides a chirally invariant
generalization of the (multi--component) Thirring model. Exploiting
the
symmetries of this model, we may reduce it in a frame where
\be
\psi_{1}^{* \a} \psi_{1}^{\a} = g_{1} ~ , ~~~~
\psi_{2}^{* \a} \psi_{2}^{\a} = g_{2},
\ee
\be
\psi_{1}^{* \a} \partial \psi_{1}^{\a} -
\psi_{1}^{\a} \partial \psi_{1}^{* \a} = -2i ~ {g_{1}}^{2} ~ , ~~~
\psi_{2}^{* \a} \bar{\partial} \psi_{2}^{\a} -
\psi_{2}^{\a} \bar{\partial} \psi_{2}^{* \a} = 2i ~ {g_{2}}^{2},
\ee
with $g_{1}$ and $g_{2}$ being constant. Here (1) and (2) denote the
upper and lower components of the Lorentz spinors $\psi^{\a}$, while
summation over the fermion species $\a = 1, ~ 2$ is implicitly
assumed.
Then, it can be verified [17] that
the composite complex fields
\be
u = {1 \over 2 ~ \sqrt{g_{1} g_{2}}} \sum_{\a=1}^{2} \bar{\psi}^{\a}
(1 + \gamma_{5}) \psi^{\a} ~ , ~~~
\bar{u} = {1 \over 2 ~ \sqrt{g_{1}
g_{2}}} \sum_{\a=1}^{2} \bar{\psi}^{\a} (1 - \gamma_{5}) \psi^{\a}
\ee
satisfy eqs.(1.4), (1.5) with
\be
g =  g_{1} g_{2}.
\ee
If there is an arbitrary coupling constant in the 4--fermion
interaction, it
will enter multiplicatively in (1.11).  Consequently, in the absence
of
the 4--fermion interaction, the unperturbed $SU(2)/U(1)$ coset model
is
recovered.

This result is
analogous to the well known relation between the Thirring
and the sine--Gordon models [18]. Note in the present case that the
chiral
invariance of the Gross--Neveu model manifests as rotational
($U(1)$ invariance) of the perturbed parafermion model. Indeed, in
the
variables (1.10), the chiral 4--fermion interaction is simply
${\mid u \mid}^{2}$. Breaking chiral invariance of the theory by
adding a
mass term $\bar{\psi}^{\a} \psi^{a}$, corresponds to introducing
the $U(1)$ violating term $ u + \bar{u}$ in the perturbed parafermion
model.
A useful consequence of our work is that the
infinite many conservation laws
of the theory (1.4) can be applied directly
to the classical $U(2)$ fermion system, using
the transformation (1.10) in the special frame (1.8), (1.9).
To the best of our
knowledge, the systematic construction of the infinite many conserved
charges of the Gross--Neveu model has not been carried out
in detail. Although its integrable properties have been studied
in general, the (off--critical) $W_{\infty}$ structure of its
currents
has not been recognized so far.

Having presented an outline of the main issues
and motivations of the present
work, we describe briefly the organization of the remaining sections.
In section 2, we examine the perturbation
of the parafermion coset by its first thermal operator, in the
framework of gauged WZW models. We also discuss the 1--soliton (and
anti--soliton) solution that exists in the large $N$ limit of the
theory.
In section 3, the physical
interpretation and geometry of the classical theory (1.4) are
subsequently
described in Minkowski space
for $g > 0, ~ g < 0$ and $g = 0$. In section 4, the infinitely many
local conservation laws are obtained by the abelianization method
of gauge connections. It turns out that for generic values of the
coupling
constant $g$, the higher spin currents of the theory are written in
terms of an off--critical generalization of the $W_{\infty}$
generators.
In section 5, connections with KdV type equations are presented. In
particular, the conserved densities of the perturbed model are
identified with the Hamiltonian densities of the (2--boson) KP
hierarchy,
thus generalizing the connection between the sine--Gordon equation
and
$SL(2)$ KdV hierarchy. Finally, in section 6, we present the
conclusions
and directions for future work. A geometrical interpretation of the
conserved charges is also given in terms of embeddings.

\section{\bf WZW Description of the Model}
\setcounter{equation}{0}
\noindent
We review first some standard results from the theory of $Z_{N}$
parafermions [19], in order to explain the form and the properties of
the
classical action (1.4). It is known that the field space of the
$SU(2)_{N}$ WZW model (see also [20, 21]) contains $N + 1$ invariant
fields
$\Phi^{(j)}$ with $j = 0, ~ 1/2, ~ 1, ~ 3/2, ~ \cdots, ~ N/2$;
$\Phi^{(0)}$
coincides with the identity operator. Each $\Phi^{(j)}$ is an $SU(2)
\times
SU(2)$ tensor with ${(2j+1)}^{2}$ components $\Phi_{m,
\bar{m}}^{(j)}$
with $m, ~ \bar{m} = -j, ~ -j+1, ~ \cdots, ~ j-1, ~ j$. These fields
have
conformal dimension $j(j+1)/(N+2)$ and $m, ~ \bar{m}$ are the $U(1)$
charges
of $\Phi_{m, \bar{m}}^{(j)}$ in the two chiral sectors of the theory.
The
principal fields $\phi_{2m, 2\bar{m}}^{(2j)}$ of the $Z_{N}$
parafermion
theory are related with the components of the WZW invariant fields by
\be
\Phi_{m, \bar{m}}^{(j)} (z, \bar{z}) = \phi_{2m, 2\bar{m}}^{(2j)} (z,
\bar{z}) ~ exp\left\{ {im \over \sqrt{N}} \chi (z) + {i \bar{m} \over
\sqrt{N}} \bar{\chi} (\bar{z}) \right\} ~ ,
\ee
where $\chi$ and $\bar{\chi}$ are the $U(1)$ bosons of the two chiral
sectors that are moded out in the construction of the
$SU(2)_{N}/U(1)$
coset model. In this notation, $\s_{k} = \phi_{k,k}^{(k)}$ are the
spin
variables and $\mu_{k} = \phi_{k,-k}^{(k)}$ are the dual spin
variables
of the parafermion theory.

We are interested in the primary field with $j = 1$, in which case
the
9 components of $\Phi^{(1)}$ can be naturally identified [20] with
the matrix elements
\be
\Phi_{ab}^{(1)} = Tr (g^{-1} T_{a} g T_{b}),
\ee
where $g$, $T$ are $SU(2)$ group elements and Lie algebra generators
respectively. The thermal operators (sometimes also called energy
operators)
are defined to be the $U(1)$ neutral fields $ \epsilon_{j} =
\phi_{0,0}^{(2j)}$ and so
\be
\epsilon_{1} = Tr (g^{-1} \s_{3} g \s_{3}).
\ee
In the large $N$ limit, the conformal dimension of all $\epsilon_{j}$
goes
to zero. An important property of the thermal operators under the
Krammers--Wannier duality $\s \leftrightarrow \mu$ is
\be
\epsilon_{j} \rightarrow (-1)^{j} ~ \epsilon_{j}.
\ee
The Krammers--Wannier duality generalizes to arbitrary
$N$ the relation between the low
and high temperature phases of the Ising model. As we will see later,
the
change of sign in $\epsilon_{1}$ under $\s \leftrightarrow \mu$
implies,
upon analytic continuation in Minkowski space,
a duality relation between the $O(4)$ non--linear $\s$--model and the
relativistic theory of vortices in a certain frame.

Consider now the Lagrangian description of the $SU(2)/U(1)$ coset
model in
terms of the $SU(2)$ gauged WZW model [3, 4, 5]. Gauging the diagonal
$U(1)$ subgroup of $SU(2)$ we obtain the action
\be
S = S_{WZW} + {N \over 2 \pi} \int Tr\left(iA \bar{\partial} g g^{-1}
-
i\bar{A} g^{-1} \partial g + A g \bar{A} g^{-1} - A \bar{A} \right),
\ee
where the gauge fields $A$, $\bar{A}$ take values in $U(1)$. The
classical
equations of motion for $A$ and $\bar{A}$ are in component form
\be
A = {1 \over 2(1-M_{33})} ~  Tr(g^{-1} \partial g ~ {\s}_{3}),
\ee
\be
\bar{A} = - ~ {1 \over 2 (1 - M_{33})} ~
Tr (\bar{\partial} g g^{-1} {\s}_{3}),
\ee
where $M_{33}$ is given by
\be
M_{33} = {1 \over 2} ~ Tr(g^{-1} {\s}_{3} g {\s}_{3}).
\ee
We fix the gauge by choosing $SU(2)$ group elements of the form
\be
g = \left( \begin{array}{ccc}
g_{0} + i g_{3} & ~ & i g_{1} \\
{}~           & ~   &  ~ \\
i g_{1} & ~ & g_{0}-ig_{3}
           \end{array}   \right),
\ee
with ${g_{0}}^{2} + {g_{1}}^{2} + {g_{3}}^{2} = 1$. Then, in this
unitary
gauge, substituting the classical equations of motion for $A$ and
$\bar{A}$ in (2.5), we obtain the action of the $SU(2)/U(1)$ coset
model
\be
S = {N \over 4 \pi} ~ \int {\partial u \bar{\partial} \bar{u} +
\partial \bar{u} \bar{\partial} u \over 1 - {\mid u \mid}^{2}}
\ee
in terms of the complex variables
\be
u = g_{0} + i g_{3} ~ , ~~~~ \bar{u} = g_{0} - i g_{3}.
\ee
Clearly, we have the condition ${\mid u \mid}^{2} \leq 1$.

The classical action (2.10) differs from the ordinary $O(3)$
non--linear
$\s$--model in that the target space metric is
${(1 - {\mid u \mid}^{2})}^{-1}$ instead of
${(1 - {\mid u \mid}^{2})}^{-2}$. As a result, the topology of the
target
space is not that of a round sphere but of two bell touching together
at
the rim ${\mid u \mid}^{2} = 1$. The action (2.10) is valid only
classically and therefore it provides a Lagrangian description of the
parafermion theory in the large $N$ limit. Quantum mechanically there
are $1/N$ corrections to the target space metric [22] and there is
also a
dilaton field from the path integral measure that insures conformal
invariance at the critical point. In this paper the quantum mechanics
of the problem will not be considered at all.

The perturbation of $S_{WZW}$ by the first thermal operator
$\epsilon_{1}$
can be gauged similarly, with no extra effort, because $\epsilon_{1}$
is the
neutral component of the primary WZW field $\Phi^{(1)}$. In the
unitary
gauge we have
\be
\epsilon_{1} = 2 ~ M_{33} = 2 ~ (2{\mid u \mid}^{2} - 1)
\ee
and so the action we obtain in the large $N$ limit of the perturbed
theory
is essentially given by eq.(1.4), where $g$ is the coupling constant
of the perturbation. In deriving (1.4) we have shifted the action by
a
constant. This adjusts the zero of the energy density and has no
effect on
the classical equations of motion.  We also note that unlike the
sine--Gordon model (1.7), the coupling constant $g$ of the
2--component
generalization (1.5), (1.6) can not be made positive always. There
are
two phases in the perturbed theory, one for $g > 0$ and one for $g <
0$,
related to each other by the Krammers--Wannier duality (2.4). The
conformal (critical) point $g = 0$ is self--dual.
Generalization of this
formalism to other pertubations is also possible.
It would be interesting to study the
Lagrangian description and the geometric interpretation  of
perturbations
driven by the higher thermal operators of parafermions, in a
systematic
way. The Lund--Regge formalism that will be adopted in the next
section
might be useful for handling the more general case as well. We hope
to
address these problems in the future.

At the classical level, the coupling constant $g$ can be normalized
to
1 or $-1$, depending on its sign, with no loss of generality.
This is possible because $g$ is
dimensionful of weight (1, 1). In Euclidean space, the two cases are
related to each other by Krammers--Wannier duality,
as it has already been pointed out. In Minkowski space,
the sign of the coupling constant can change by interchanging the
role of
space and time coordinates. From now on we
consider the theory (1.4) defined in Minkowski space and study its
physical interpretation for $g > 0$, $g < 0$ and $g = 0$ in the
light--cone
coordinates
\be
\partial = \partial_{\s} + \partial_{\tau} ~ , ~~~~
\bar{\partial} = \partial_{\s} - \partial_{\tau}.
\ee
The space and time variables will be $\s$ and $\tau$ in all cases.

An important property of the perturbed parafermion
theory is that it admits soliton
solutions, in analogy with the sine--Gordon model. To make the
comparison
easier, we introduce the variables $\lambda$ and $\theta$
\be
u = cos \theta ~ e^{i \lambda} ~ , ~~~~ \bar{u} = cos \theta ~
e^{-i \lambda}.
\ee
In the static limit $\partial = \bar{\partial}$,
the 1--soliton solution of the classical
equations of motion (1.5), (1.6) was found long time ago [12]. For
$g > 0$ (and normalized to 1) the 1--soliton is
\be
\theta(z) = sin^{-1} \left\{ {\sqrt{1- A^{2}} \over cosh
\left( z ~ \sqrt{1 - A^{2}}
\right)} \right\} = sin^{-1} \left\{ \sqrt{1 - A^{2}} sin \left(
2 ~ tan^{-1} exp(z \sqrt{1 - A^{2}}) \right) \right\} ,
\ee
\be
\lambda(z) = A \int^{z} d z^{\prime} ~
{tan}^{2} \theta(z^{\prime}),
\ee
where $A$ is constant. In the limit $A \rightarrow 0$  we have
$\lambda = 0$ and the usual sine--Gordon 1--soliton solution
$\theta = 2 ~ tan^{-1} expz$ is obtained (see for instance [23]).
The corresponding anti--soliton solution is obtained by shifting
$\theta \rightarrow \pi - \theta$.
As we will see later, the parameter $A$ is the
$U(1)$ charge of the solution and so the sine--Gordon model describes
the zero charge sector of the theory.
A localized lump travelling
with constant velocity is
obtained, as usual, by Lorentz  transformation.

Multisolitons solutions can also be
constructed, using standard techniques from soliton theory.
The difference between $g > 0$ and $g < 0$ in the soliton solutions
has been considered in ref.[24], together with some related issues.
An analogous solution, for $g = 0$, was found by Bardacki et.al.[4].
In the static limit of the theory,
\be
\theta (z) = {sin}^{-1} \left\{ \sqrt{1 - A^{2}} ~ sin z \right\}
\ee
and $\lambda (z)$ also given by eq.(2.16), solve the classical
equations
of motion without the thermal perturbation.

The existence of soliton solutions in the perturbed parafermion model
might have important consequences for the associated scattering
problem.
This sector has not been included so far  in determining the
$S$--matrix of the theory\footnote{I thank T. Hollowood for some
discussions on this point.}.
Naive consideration of the large $N$ limit
shows that $S = 1$ for this model [6, 25]. It is not clear at this
point
whether the soliton
solutions persist for finite $N$ or whether they are characteristic
of the large $N$ limit. It might be possible to address this question
by considering the $1/N$ corrections to the classical action (1.4)
from
the gauged WZW point of view. In any event, the scattering of these
solitons should be investigated separately in the future.

\section{\bf Geometry of the Thermal Perturbation}
\setcounter{equation}{0}
\noindent
In this section we describe geometrically the classical equations of
motion of the perturbed parafermion model. The main subject
of the section is the
embedding of a 2--dim surface $S$ with local coordinates $\s$, $\tau$
and metric
\be
{ds}^{2} = {cos}^{2} \theta {d \s}^{2} + {sin}^{2} \theta {d
\tau}^{2}
\ee
in a 3--dim space of constant curvature, which in turn is embedded in
4--dim flat space.
The differential equations that
the metric\footnote{Although we use the same symbol $g$ for the
metric
and the coupling constant, the distinction between the two will be
obvious
from the context they appear.}  $g$
and the extrinsic curvature tensor $K$ of $S$ have to
satisfy, in order to have a solution to the embedding problem, are
given
by the Gauss--Codazzi integrability conditions (see for instance
[26]).
These conditions admit a field theoretic interpretation which
helps us to understand the physics
and geometry of the first thermal perturbation of the parafermion
coset.
The technical details of the embedding will be  considered
later and they are intimately related with the physical
interpretation of
the perturbed coset model.  The cases $g < 0$ and $g > 0$ are treated
separately, since the results we obtain depend crucially on the sign
of
the coupling constant. In this framework we also revisit the critical
point $g = 0$ and find that the $SU(2)/U(1)$ coset model
provides an effective description of the transverse modes of the
4--dim
Nambu--Goto string in the orthonormal gauge.

The formalism we adopt here originates in the work by Lund and Regge
[12, 13] (see also [14]), but since it is rather unknown we review
briefly the main ideas applicable to our model.
The considerations are
entirely local, with no proper reference to boundary conditions.
Throughout this section, the perturbed parafermion model is defined
on
2--dim Minkowski space. Analytic continuation
to Minkowski space is neccessary,
in order to describe the physics and geometry of the theory via
embeddings.

(i) \underline{$g < 0$}: In this case, the perturbed model (1.4)
is classically equivalent to the
2--dim $O(4)$ non--linear $\s$--model, reduced in a certain frame.
Recall that the $O(4)$ model consists of four scalar fields
$\eta_{i}(\s, \tau)$ ($i = 1, ~ 2, ~ 3, ~ 4$)
interacting through the quadratic
constraint
\be
{(\eta_{i})}^{2} = 1.
\ee
The solutions of the classical equations of motion
\be
({\partial_{\s}}^{2} - {\partial_{\tau}}^{2}) \eta_{i} +
\left( {(\partial_{\s} \eta_{k})}^{2} - {(\partial_{\tau}
\eta_{k})}^{2}
\right) \eta_{i} = 0,
\ee
describe 2--dim surfaces (parametrized
by $\s$ and $\tau$) embedded in the 3--dim sphere (3.2), which in
turn is
embedded in flat 4--dim Euclidean space. The embedding variables are
simply
$\eta_{i}$. It is natural to reduce the $O(4)$ non--linear
$\s$--model
in the frame
\be
{(\partial_{\s} \eta_{i})}^{2} + {(\partial_{\tau} \eta_{i})}^{2} = 1
{}~ ,
{}~~~~ (\partial_{\s} \eta_{i}) (\partial_{\tau} \eta_{i}) = 0
\ee
by exploiting the invariances of the classical theory. When the
conditions (3.4) are satisfied, the induced metric on the 2--dim
embedded
surface is given by eq.(3.1), where
\be
{cos}^{2} \theta = {(\partial_{\s} \eta_{i})}^{2} ~ , ~~~~
{sin}^{2} \theta = {(\partial_{\tau} \eta_{i})}^{2}.
\ee
Note that the $O(4)$ model is defined in Minkowski space, while the
associated embedding problem is entirely Euclidean.

The formulation of an embedding problem requires apart from the
metric $g$,
knowledge of the extrinsic curvature tensor $K$. Its components are
defined to be
\be
K_{\s \s} = ({\partial_{\s}}^{2} \eta_{i}) Z_{i}^{(3)} ~ , ~~
K_{\tau \tau} = ({\partial_{\tau}}^{2} \eta_{i}) Z_{i}^{(3)} ~ , ~~
K_{\s \tau} = K_{\tau \s} = (\partial_{\s} \partial_{\tau} \eta_{i})
Z_{i}^{(3)},
\ee
where $Z^{(3)} = Z^{(1)} \times Z^{(2)}$ with
\be
Z_{i}^{(1)} = {1 \over cos \theta} (\partial_{\s} \eta_{i}) ~ , ~~~~
Z_{i}^{(2)} = {1 \over sin \theta} (\partial_{\tau} \eta_{i}).
\ee
We also choose $Z_{i}^{(4)} = \eta_{i}$ to complete an orthonormal
tetrad
in 4--dim Euclidean space, where the embedding takes place.
The classical equations of motion of the reduced $O(4)$ non--linear
$\s$--model are equivalent in the special frame (3.4) to
\be
K_{\s \s} = K_{\tau \tau}.
\ee
This is obtained by multiplying the classical equations of motion
(3.3)
with $Z_{i}^{(3)}$. Multiplication with $Z_{i}^{(1)}$, $Z_{i}^{(2)}$
and $Z_{i}^{(4)}$ leads to identities
in the special frame (3.4). The reduction of the $O(4)$ model in this
frame is analogous to the reduction of the chirally invariant $U(2)$
Gross--Neveu model, described in the introduction.

It is clear that the classical equations of motion of the reduced
theory
can be formulated as an embedding
problem in the 3--dim sphere (3.2), with the extrinsic curvature
satisfying
the physical requirement (3.8). This embedding is possible provided
that
the Gauss--Codazzi conditions on $g$ and $K$ are satisfied. In the
present case they assume the form
\be
\partial_{\tau} (tan \theta ~ K_{\s \s}) = \partial_{\s} (tan \theta
{}~
K_{\s \tau}) ~ , ~~~~ \partial_{\s} (cot \theta ~ K_{\s \s}) =
\partial_{\tau} (cot \theta ~ K_{\s \tau}),
\ee
\be
R = {(g^{\a \beta} K_{\a \beta})}^{2} -
g^{\a \gamma} g^{\beta \delta} K_{\a \beta}
K_{\gamma \delta} + 2,
\ee
where $R$ is the scalar curvature of the 2--dim metric (3.1)
\be
R = {2 \over sin \theta \cos \theta} ({\partial_{\tau}}^{2} \theta -
{\partial_{\s}}^{2} \theta).
\ee
The constant term in eq.(3.10) is the curvature contribution of
the 3--dim sphere (3.2) where the embedding takes place.
The first condition in eq.(3.9) implies the existence of a scalar
field
$\lambda$ such that $K_{\s \s} = cot \theta ~ \partial_{\s} \lambda$
and
$K_{\s \tau} = cot \theta ~ \partial_{\tau} \lambda$. Introducing
complex variables $u, \bar{u}$ as in eq.(2.13) and light--cone
coordinates
as in eq.(2.16), we obtain
\be
K_{\s \s} + K_{\s \tau} = i ~ { u \partial \bar{u} - \bar{u} \partial
u
\over 2 ~ \mid u \mid \sqrt{1 - {\mid u \mid}^{2}}} ~ , ~~~~
K_{\s \s} - K_{\s \tau} = i ~ { u \bar{\partial} \bar{u} - \bar{u}
\bar{\partial} u \over 2 ~ \mid u \mid \sqrt{1 - {\mid u \mid}^{2}}}
{}~ .
\ee
Then, it is straightforward to verify that the remaining
Gauss--Codazzi
conditions become the classical equations of motion of the
perturbed parafermion model in Minkowski space, with $g = -1$.

The reconstruction of $S$ (and hence ${\eta}_{i}$) can be done by
solving the Gauss--Weingarten equations for the orthonormal vectors
$Z^{(1)}$, $Z^{(2)}$, $Z^{(3)}$ and $Z^{(4)}$. These are first order
linear differential equations, forming a system, whose compatibility
conditions are provided by the Gauss--Codazzi equations (for details
see [26]).
The classical equivalence between the two theories, both
defined in 2--dim Minkowski space,
has been established with the aid of a Euclidean
embedding problem. In this case, the curvature of the 3--dim sphere
(3.2) determines the coupling constant (it
turns out to be $-2 g$) and hence $g$ is negative.

The physical interpetation of the perturbed theory with $g > 0$
will be addressed differently, without
interchanging $\s \leftrightarrow \tau$.
Following Lund and Regge [12], we find that
the phase of the theory with $g > 0$ describes (in a certain gauge)
the
relativistic motion of vortices in a constant external field. We also
adopt this picture in order to show that the transverse modes of the
4--dim Nambu--Goto string in the orthonormal gauge are effectively
described by the conformal limit of the $SU(2)/U(1)$ coset model.

(ii) \underline{$g > 0$}: Consider first the string action
in 4--dim Minkowski space
\be
S = -N \int \sqrt{-detg}~ d \s d \tau + f \int A_{\mu \nu} (X)
\partial_{\s} X^{\mu} \partial_{\tau} X^{\nu} d \s d \tau -
{1 \over 4} \int F_{\mu} F^{\mu} d^{4} y,
\ee
where $\s$, $\tau$ are local coordinates on the string world--sheet
$\S$ and $det g$ is the determinant of the 2--dim metric on $\S$. The
first term is the usual Nambu--Goto action, while the second term
represents a self--interaction of Kalb--Ramond type [27]. $A_{\mu
\nu}(X)$
is an anti-symmetric tensor (axionic background) and $F_{\mu}$ is
defined to be
\be
F^{\mu} = {1 \over 2}~ {\epsilon}^{\mu \nu \lambda \rho}
\partial_{\rho}
A_{\nu \lambda}.
\ee
$X^{\mu} (\s, \tau)$ ($\mu = 0, ~ 1, ~ 2, ~ 3$) are the embedding
variables
of the string in 4--dim Minkowski space with signature $-+++$. When
the
spatial components of $A_{\mu \nu}$ are linear in $X_{i}$
($i = 1, ~ 2, ~ 3$), ie
\be
A_{ij} = \epsilon_{ijk} X^{k},
\ee
this action describes the relativistic motion of vortices (strings)
in a
superfluid [12] (see also [28] for the non--relativistic limit).
In this case, $F^{i}$ is identified with the velocity of the fluid
\be
F^{i} = \epsilon^{ijk} \partial_{j} A_{k0} = v^{i}
\ee
and the last term in eq.(3.13) represents the hydrodynamic action of
the
superfluid.

Following Lund and Regge we study the relativistic motion of vortices
in a
uniform static external field. We choose a Lorentz frame in which
\be
X^{0} = \tau ~, ~~~~ F^{i} = 0
\ee
and $F^{0}$ is constant and introduce the effective coupling constant
of
the theory
\be
c = {f F^{0} \over 2 N}.
\ee
At this point we do not normalize $c$ to 1, keeping its dependence
on $f$ and $N$ explicitly. We need this in order to understand the
physical
interpretation of the limit $c \rightarrow 0$
that will be considered later.
We also choose the orthonormal gauge for the induced metric on the
vortex (string) world--sheet, which leads to the quadratic
constraints
\be
{(\partial_{\s} X_{i})}^{2} + {(\partial_{\tau} X_{i})}^{2} = 1 ~ ,
{}~~~~
(\partial_{\s} X_{i}) (\partial_{\tau} X_{i}) = 0
\ee
in the Lorentz frame (3.17). This gauge choice is analogous to the
reduction
of the $O(4)$ non--linear $\s$--model (3.4) introduced earlier. Then,
the
vortex dynamics is determined entirely by the classical equations of
motion
\be
({\partial_{\s}}^{2} - {\partial_{\tau}}^{2}) X_{i} + 2 ~ c \left(
({\partial}_{\s} \vec{X}) \times ({\partial}_{\tau} \vec{X})
\right)_{i} = 0
\ee
plus the constraints (3.19).

Note that for uniform static external field, the last term in the
action (3.13) can be dropped out (it is just a number). The theory in
this
case describes a 4--dim bosonic string in
an axionic background of vortex type.
For $f = 0$, ie $c = 0$, it describes the propagation of a free
bosonic
string in 4--dim Minkowski space in the orthonormal gauge with $X_{0}
=
\tau$. It is important to realize that the Lund--Regge model for
vortex
dynamics in a uniform static external field is classically equivalent
to
the perturbed parafermion coset (1.4) with $g = c^{2} > 0$, upon
analytic continuation in Minkowski space. This formalism
is certainly unphysical for $g < 0$, since the coupling
constant of the vortex self--interaction term will be imaginary in
that
case. We also point out that the results presented in the sequel are
independent of the Lorentz frame $X_{0} = \tau$, up to Lorentz
transformations
in the ($\s, \tau$) space.

To illustrate the vortex--like description of the perturbed
parafermion model
with $g > 0$ it is convenient to incorporate the classical equations
of
motion (3.20) in the integrability conditions of an embedding
problem, in
analogy with the $g < 0$ case. Let $S$ be the projection of the
string
world--sheet $\S$ in the $X_{0} = \tau$ hyperplane\footnote{To know
$\vec{X},$ it is sufficient to consider $S$ for $X^{0} = \tau$.}.
$S$ is a Euclidean
surface, which in the orthonormal frame has an induced metric given
by
eq.(3.1). We have
\be
{(\partial_{\s}X_{i})}^{2} = {cos}^{2} \theta ~ , ~~~~
{(\partial_{\tau}X_{i})}^{2} = {sin}^{2} \theta.
\ee
In analogy with the $g < 0$ case, we also consider the extrinsic
curvature
tensor of $S$,
\be
K_{\s \s} = ({\partial_{\s}}^{2} X_{i}) Z_{i}^{(3)} ~ , ~~
K_{\tau \tau} = ({\partial_{\tau}}^{2} X_{i}) Z_{i}^{(3)} ~ , ~~
K_{\s \tau} = K_{\tau \s} = (\partial_{\s} \partial_{\tau} X_{i})
Z_{i}^{(3)}
\ee
with $Z^{(3)} = Z^{(1)} \times Z^{(2)}$, where
\be
Z_{i}^{(1)} = {1 \over cos \theta} \partial_{\s} X_{i} ~ , ~~~~
Z_{i}^{(2)} = {1 \over sin \theta} \partial_{\tau} X_{i}.
\ee
In terms of these variables, the classical
evolution of vortices in the
orthonormal frame is entirely determined by the embedding of $S$ in
the 3--dim Euclidean space $X_{0} = \tau$, which in turn is embedded
in
4--dim Minkowski space. The components of the extrinsic curvature
tensor
have to satisfy the condition
\be
K_{\s \s} - K_{\tau \tau} + 2c ~ sin \theta cos \theta = 0,
\ee
which follows from the classical equations of motion (3.20) by
multiplication with $Z_{i}^{(3)}$. Multiplication with $Z^{(1)}$ and
$Z^{(2)}$ leads to identities in the special frame (3.19). Hence,
eq.(3.24) encodes all the information contained in eq.(3.20).

The embedding problem here is different in that the
3--dim space where the embedding takes place
is flat and the extrinsic curvature satisfies the condition
(3.24) rather than (3.8). The Gauss--Codazzi
integrability conditions have to be satisfied, however,
in order to have a solution to the problem. In this case they read
\be
\partial_{\tau} \left( tan \theta (K_{\s \s} + K_{\tau \tau}) \right)
=
2 \partial_{\s} (tan \theta K_{\s \tau}) ~ , ~~~
\partial_{\s} \left( cot \theta (K_{\s \s} + K_{\tau \tau}) \right) =
2 \partial_{\tau} (cot \theta K_{\s \tau}),
\ee
\be
R = {(g^{\a \beta} K_{\a \beta})}^{2} - g^{\a \gamma} g^{\beta
\delta}
K_{\a \beta} K_{\gamma \delta},
\ee
where $R$ is the curvature of the 2--dim metric (3.1),
given again by eq.(3.11).
Introducing light--cone variables $z$, $\bar{z}$ and complex
coordinates $u$, $\bar{u}$ as before, we find that the first
condition
in eq.(3.25) implies
\be
K_{\s \s} + K_{\tau \tau} + 2 ~ K_{\s \tau} = i ~
{u \partial \bar{u} - \bar{u} \partial u \over \mid u \mid
\sqrt{1 - {\mid u \mid}^{2}}} ~ , ~~~
K_{\s \s} + K_{\tau \tau} - 2 ~ K_{\s \tau} = i ~
{u \bar{\partial} \bar{u} - \bar{u} \bar{\partial} u \over \mid u
\mid
\sqrt{1 - {\mid u \mid}^{2}}} ~ .
\ee
It is staightforward to verify using eq.(3.24) that the remaining
Gauss--Codazzi integrability conditions become the
classical equations of motion of the perturbed parafermion model in
Minkowski space, with $g = c^{2}$. The reconstruction of $S$ (and
hence $X_{i}$) can be done by solving the corresponding
Gauss--Weingarten equations for the orthonormal vectors $Z^{(1)}$,
$Z^{(2)}$ and $Z^{(3)}$.

(iii) \underline{$g = 0$}: When the coupling constant $f$ of the
antisymmetric tensor $A_{\mu \nu}$ is zero, the theory reduces to
the Nambu--Goto string propagating in 4--dim Minkowski space.
As it has been pointed out already,
in the orthonormal gauge with $ X_{0} = \tau$,
the unperturbed $SU(2)/U(1)$ coset model, when it is defined in
Minkowski space, describes the classical dynamics of the transverse
modes of a free string. This result is very intriguing and probably
it can be generalized to more arbitrary backgrounds. In a separate
publication [15], the implications of this equivalence are analyzed
in
detail. The infinitely many conservation laws of the parafermion
coset model
can be easily applied to string dynamics to yield hidden symmetries
(Backlund transformations) in their classical solution space. In the
next
section we analyze the structure of the conserved charges for
arbitrary values of the coupling constant $g$ and establish their
relation with $W_{\infty}$. Then, a geometric interpretation of the
higher
spin currents follows immediately from the embedding problem that
describes the classical equations of motion of the theory.

The classical equivalence between 4--dim string theory and the
perturbed
parafermion model with $g \geq 0$ is not restricted to the special
frame $X_{0} = \tau$. For $X_{0} = \tilde{\tau}$, where
\be
\tilde{\tau} = sinh \a ~ \s + cosh \a ~ \tau ~, ~~~
\tilde{\s} = cosh \a ~ \s + sinh \a ~ \tau
\ee
are related by Lorentz transformation in $(\s, \tau)$ space, the
embedding
problem remains essentially the same. The projection of the string
world--sheet is performed now in the
$X_{0} = \tilde{\tau}$ 3--space, but the
classical equations of motion that follow from the Gauss--Codazzi
conditions remain unchanged [12]. For this reason, the equivalence
between
the two theories is independent of the Lorentz frame. This issue does
not arise for $g < 0$, because the physics of the problem does not
require
the choice of a Lorentz frame to formulate the corresponding
embedding.

The soliton solution (2.14)--(2.15) of the classical
equations of motion with $g > 0$ becomes relevant for string
propagation in an axionic background
of vortex type. It would be interesting to study further the
localized
properties of this soliton solution, directly in the string
variables $X_{i}$. The issue of boundary conditions and the physical
interpretation of coset model solutions has to be
addressed properly in the future, in the context of string theory.

\section{\bf Local Conservation Laws}
\setcounter{equation}{0}
\noindent
We turn now to the explicit construction of the infinitely many
conservation laws of the perturbed parafermion theory
for arbitrary values of the coupling constant $g$. For this, it
is convenient to rewrite the classical equations of motion (1.5),
(1.6)
as a zero curvature condition. We introduce the linear system of
differential
equations
\be
\partial \Phi = A ~ \Phi ~ , ~~~~
\bar{\partial} \Phi = B ~ \Phi,
\ee
where $A$ and $B$ are $ 2 \times 2 $ matrices depending on $z$,
$\bar{z}$
and a spectral parameter $\lambda$
\be
A = A_{0} + \lambda E ~ , ~~~~ B = B_{0} + {\lambda}^{-1} B_{1} .
\ee
The spectral parameter should not be confused with the scalar field
$\lambda$ appearing in the definition (2.13) of $u$ and $\bar{u}$.
$ A_{0} $, $ B_{0} $ and $ B_{1} $ are taken to be
\be
A_{0} = - {1 \over 4 ~ {\mid u \mid}^{2}
(1 - {\mid u \mid}^{2})} ~
\left( \begin{array}{cc}
(2{\mid u \mid}^{2} - 1)(u \partial \bar{u} - \bar{u} \partial u) &
4i \mid u \mid \sqrt{1 - {\mid u \mid}^{2}} ~ \bar{u} \partial u \\
{}~ & ~ \\
4i \mid u \mid \sqrt{1 - {\mid u \mid}^{2}} ~ u \partial \bar{u}  &
-(2{\mid u \mid}^{2} - 1)(u \partial \bar{u} - \bar{u} \partial u) \\
       \end{array} \right),
\ee
\ba
B_{0} & = & {u \bar{\partial} \bar{u} - \bar{u}
\bar{\partial} u \over 4 ~ {\mid u \mid}^{2} (1 - {\mid u \mid}^{2})}
{}~ E,\\
B_{1} & = &
{g \over 4} ~
\left( \begin{array}{cc}
2{\mid u \mid}^{2} - 1       & -2i \mid u \mid \sqrt{1 - {\mid u
\mid}^{2}} \\
{}~                            &  ~ \\
2i \mid u \mid \sqrt{1 - {\mid u \mid}^{2}} & -(2{\mid u \mid}^{2} -
1) \\
       \end{array} \right)
\ea
and $ E $ is the constant matrix
\be
E = \left( \begin{array}{ccc}
1 & ~ & 0 \\
{}~ & ~ & ~ \\
0 & ~ & -1 \\
           \end{array} \right).
\ee
Then, it may be easily verified that the  compatibitity (zero
curvature)
condition of the linear system (4.1),
\be
[ \partial - A ~ , ~ \bar{\partial} - B ] = 0,
\ee
is equivalent to the classical equations of motion (1.5), (1.6),
for all values of $\lambda$.

The physical meaning of the spectral parameter $\lambda$ can be
understood in terms of the Lorentz transformation (3.28). Suppose
that
the linear system (4.1) is given initially with $\lambda = 1$. The
Lorentz transformation $(\s , \tau) \rightarrow
(\tilde{\s} , \tilde{\tau})$ amounts to the rescaling
$\partial \rightarrow exp(- \a) ~ \partial$, $\bar{\partial}
\rightarrow exp \a ~ \bar{\partial}$ and $A_{0} \rightarrow exp(- \a)
{}~
A_{0}$, $ B_{0} \rightarrow exp \a ~ B_{0}$. Absorbing the Lorentz
factor, is equivalent to choosing $\lambda = exp \a$ in the gauge
connections (4.2). Consequently, the independence of the classical
equations of motion on $\lambda$ is really a statement about Lorentz
invariance.
As we will see shortly, the
presence of a spectral parameter in the linear system (4.1) is
neccessary
to derive the infinitely many conservation
laws of the theory in a systematic
way. The results are valid equally well in Euclidean and Minkowski
space.

Consider the static limit $ \partial = \bar{\partial}$ first. It
follows
immediately from $[ \partial - A ~ , ~ \partial - B] = 0 $ that the
quantities
\be
I_{n} = {1 \over 2} ~ Tr {(A-B)}^{n} ~ , ~~~~ n = 1, ~ 2 , ~ 3, ~
\cdots ~ ,
\ee
are conserved, ie $\partial I_{n} = 0$. This is a well known result
for
1--dim integrable systems. If the number of degrees of freedom is
finite,
the quantities $I_{n}$ will not be all independent.
For the present model we find that only $ I_{2} $
is independent, because
\be
I_{2k-1} = 0 ~ , ~~~I_{2k} = {(I_{2})}^{k} ~ , ~~~ k = 1  , ~ 2  ,
{}~ 3 , \cdots ~ .
\ee
Explicit calculation shows that
\be
I_{2} = {1 \over 4} ~ Q^2 - H - \lambda Q + {g \over 4 ~ \lambda} ~ Q
+ {1 \over 2} ~ g + {\lambda}^{2} + {g^{2} \over 16 ~ {\lambda}^{2}},
\ee
where
\be
H = {\partial u \partial \bar{u} \over 1 - {\mid u \mid}^{2}} +
g ~ {\mid u \mid}^{2} ~ , ~~~ Q = {u \partial \bar{u} - \bar{u}
\partial u
\over 1 - {\mid u \mid}^{2}}.
\ee
Since $ \partial I_{2} = 0 $ for all values of $\lambda$, we arrive
at the
conservation laws
\be
\partial H = 0 = \partial Q
\ee
in the static limit of the theory. $H$ is the energy and $Q$ is the
$U(1)$
charge of a given configuration $u$, $\bar{u}$. For the 1--soliton
(2.14),
(2.15), with $g$ normalized to 1, we find that $H = 1$ and $Q = 2A$,
provided that the charge of the theory is made real by multiplication
with
$i$. This gives a physical interpretation to the arbitrary parameter
$A$
that determines the soliton solution. Solitons with zero $U(1)$
charge are
simply sine--Gordon solitons.

In the general case, the
conserved quantities of the theory are not given by eq.(4.8) and
a field theoretical
prescription is required to obtained the currents associated
with the 2--dim zero curvature condition (4.7). The abelianization
method of gauge connections [8, 9] provides the algorithm for their
systematic construction.
The main idea is to introduce a family of $\lambda$--dependent gauge
transformations
\be
T = 1 + \sum_{i=1}^{\infty} {\lambda}^{-i} t_{i},
\ee
where $t_{i}$ depend on $z$, $\bar{z}$ and take values in $GL(2)$
with
$det ~ T \neq 0$. Then, it is always possible to choose $\{t_{i}\}$
appropriately so that the gauge transformed connections
\be
T^{-1}(\partial - A_{0} - \lambda E) ~ T \equiv \partial +
\sum_{j=0}^{\infty} {\lambda}^{-j} \tilde{A}_{j} - \lambda E ,
\ee
\be
T^{-1}(\bar{\partial} - B_{0} - {\lambda}^{-1} B_{1}) ~ T \equiv
\bar{\partial} - \sum_{j=0}^{\infty} {\lambda}^{-j} \tilde{B}_{j}
\ee
commute with the constant matrix $E$ (4.6), ie
\be
[\tilde{A}_{j} ~ , ~ E] = 0 = [\tilde{B}_{j} ~ , ~ E]
\ee
for all values of $ j $. Since $E$ is not the identity matrix,
$\{\tilde{A}_{j}\}$ and $\{\tilde{B}_{j}\}$ commute among themselves
and the zero curvature condition (4.7) yields upon gauge
transformation
an infinite number of functionally independent non--trivial
conservation
laws
\be
\bar{\partial} \tilde{A}_{j} + \partial \tilde{B}_{j} = 0 ~ , ~~~ j =
0, ~
1, ~ 2, ~ \cdots ~ ,
\ee
one to each order in the $1/{\lambda}$ expansion.
The presence of a spectral parameter in the zero curvature
formulation of the problem is indeed essential.
The validity of the equations for all values of $\lambda$ implies
order by
order the infinite hierarchy of conservation laws (4.17). For $E$
diagonal,
as it is the case here, the abelianization method amounts to the
diagonalization of the gauge connections $A$ and $B$. It is important
to
emphasize that the diagonalization of $B$ is not automatic and it
can be achieved only on--shell.

Having presented the essential ingredients of this method,
we may proceed with explicit calculations. Note first that the
diagonalization of $A$ implies an infinite set of conditions. We have
\be
\tilde{A}_{0} + [E ~ , ~ t_{1}] + A_{0} = 0
\ee
to order ${\lambda}^{0}$, while for $ j > 0 $ we
obtain the recursive relations
\be
\tilde{A}_{j} + [E ~ , ~ t_{j+1}] + A_{0} t_{j} - \partial t_{j} +
\sum_{k=0}^{j-1} t_{j-k}  \tilde{A}_{k} = 0,
\ee
one to each order in ${\lambda}^{-j}$. Since all $\tilde{A}_{j}$ have
to
commute with $E$, their most general form is
\be
\tilde{A}_{j} = \left( \begin{array}{ccc}
h_{j}^{(1)}   & ~      & 0 \\
{}~             & ~      & ~ \\
0             & ~      & h_{j}^{(2)} \\
                       \end{array} \right) ~ , ~~~ j \geq 0.
\ee
The problem now is to solve these recursive relations, in order to
determine
the form of $\tilde{A}_{j}$ and the gauge transformation $T$.

Note that the solution to the diagonalization
problem of $A$ is not uniquely determined.
There is the freedom to set the diagonal elements of all
$t_{i}$ equal to zero, with no loss of generality. Indeed, any $2
\times 2$
matrix can be written in the form $X + [E, Y]$, where $X$ is a
diagonal
matrix and $Y$ off--diagonal.
Taking this decomposition  into account
in the $\tilde{A}_{j} + [E, t_{j+1}]$ part of the recursive relations
(4.18), (4.19), it can be easily seen that for diagonal
$\tilde{A}_{j}$,
as they should be,  $t_{j+1}$ can be chosen so that their diagonal
elements are zero.  This gauge choice fixes the form of $t_{i}$ and
$\tilde{A}_{j}$ uniquely and the solution that results from the
recursive
relations has no free parameters; everything is functionally
dependent
on the matrix elements of $A_{0}$. This gauge choice will be assumed
from now on.

Gauge transformations with
non--zero diagonal elements, modify the solution for $\tilde{A}_{j}$
by
total $\partial$--derivative terms, via the recursive relations
(4.19). This modification is supplemented in $\tilde{B}_{j}$ by
substracting the $\bar{\partial}$--derivative of these additional
terms.
This has no effect on the local conservation laws (4.17)
and it reflects the freedom we have in writing down the corresponding
currents. We also note that the
gauge transformation $T$ is not neccessarily restricted
to $SL(2)$, but it can take values in $GL(2)$.
In the gauge where the diagonal elements of all $t_{i}$ are
set equal to zero, the determinant of $T$ is
\be
det T = 1 - t_{1}^{12} t_{1}^{21} {\lambda}^{-2} -
(t_{1}^{12} t_{2}^{21} + t_{1}^{21} t_{2}^{12}) {\lambda}^{-3} +
\cdots ~ .
\ee
Therefore, it is natural to expect that for $j \geq 2$ the matrices
$\tilde{A}_{j}$ will not be traceless. We could have chosen a gauge
(order by order in $1/ \lambda$) so that $det T = 1$, to preserve the
traceless condition on all gauge connections. However, the two
gauges are related to each other by the $U(1)$ element of $GL(2)$,
which is proportional to the $2 \times 2$ unit matrix,
\be
1 + {1 \over 2} ~ t_{1}^{12} t_{1}^{21} {\lambda}^{-2} +
{1 \over 2} ~ (t_{1}^{12} t_{2}^{21} + t_{1}^{21}
t_{2}^{12}) {\lambda}^{-3} + \cdots
\ee
and the difference on the diagonalized gauge connections $\tilde{A}$,
$\tilde{B}$ is therefore a total derivative trace term. We choose
to work with a gauge in $GL(2)$, rather than $SL(2)$,
in order to simplify the form of the recursive relations.

{}From eq.(4.18) we have immediately
\be
h_{0}^{(1)} = - A_{0}^{11} = - h_{0}^{(2)}
\ee
and
\be
t_{1}^{12} = - {1 \over 2} A_{0}^{12} ~ , ~~~~  t_{1}^{21} =
{1 \over 2} A_{0}^{21},
\ee
in terms of the matrix elements of $A_{0}$.
This is the initial data for the recursive relations with $j > 0$.
To bring the gauge connection $A$ into the desired form, the
following system of equations has to be iterated:
\be
h_{j}^{(1)} = - A_{0}^{12} t_{j}^{21} ~ , ~~~~
h_{j}^{(2)} = - A_{0}^{21} t_{j}^{12},
\ee
\be
2 ~ t_{j+1}^{12} = - A_{0}^{11} t_{j}^{12} + \partial t_{j}^{12} -
\sum_{k=0}^{j-1} h_{k}^{(2)} t_{j-k}^{12},
\ee
\be
2 ~ t_{j+1}^{21} = - A_{0}^{11} t_{j}^{21} - \partial t_{j}^{21} +
\sum_{k=0}^{j-1} h_{k}^{(1)} t_{j-k}^{21}.
\ee
These relations follow from eq.(4.19), using the diagonal form (4.20)
for
$\tilde{A}_{j}$ and the off--diagonal gauge for $t_{j}$.

The solution (4.23) for the matrix elements of $\tilde{A}_{0}$
can be written in terms of $u$ and $\bar{u}$ as
\be
h_{0}^{(1)} + h_{0}^{(2)} = 0 ~ , ~~~
h_{0}^{(1)} - h_{0}^{(2)} = {2 {\mid u \mid}^{2} - 1 \over 2 ~
{\mid u \mid}^{2} (1 - {\mid u \mid}^{2})} (u \partial \bar{u} -
\bar{u} \partial u).
\ee
On the other hand, since $\tilde{B}_{0} = B_{0}$, we derive to this
order
the non--chiral conservation law
\be
\bar{\partial} J + \partial \bar{J} = 0,
\ee
where
\be
J = {u \partial \bar{u} - \bar{u} \partial u \over 1 - {\mid u
\mid}^{2}}
= 2 (h_{0}^{(1)} - h_{0}^{(2)}) - \partial \left( log ~ {u \over
\bar{u}}
\right),
\ee
\be
\bar{J} = {u \bar{\partial} \bar{u} - \bar{u} \bar{\partial} u \over
1 - {\mid u \mid}^{2}} = 2 (B_{0}^{11} - B_{0}^{22}) + \bar{\partial}
\left( log ~ {u \over \bar{u}} \right).
\ee
The ${\lambda}^{0}$ conserved currents are written conveniently in
this form, in order to identify $J$ and $\bar{J}$ with the two
components
of the $U(1)$ current associated with the symmetry of the action
(1.4)
under $ u \rightarrow u e^{i \epsilon}$ and $ \bar{u}
\rightarrow \bar{u} e^{-i \epsilon}$. Note that the conservation
law (4.29) is independent of the coupling constant $g$, simply
because the
term ${\mid u \mid}^{2}$ in the potential of the action
is invariant under $U(1)$ rotations.
In the static limit $\partial = \bar{\partial}$, we have $J = \bar{J}
=
Q$ and the conservation law $\partial Q = 0$ is recovered.

Using eqs.(4.24) and (4.25), we obtain the following
expression for $\tilde{A}_{1}$ in terms of $u$ and $\bar{u}$:
\be
h_{1}^{(1)} + h_{1}^{(2)} = 0 ~ , ~~~~  h_{1}^{(1)} - h_{1}^{(2)} =
{\partial u \partial \bar{u} \over 1 - {\mid u \mid}^{2}}.
\ee
Explicit calculation also shows that
\be
\tilde{B}_{1} = B_{1} + [B_{0} ~ , ~ t_{1}] - \bar{\partial} t_{1} =
{1 \over 4} ~ g ~ (2 {\mid u \mid}^{2} - 1) ~ E,
\ee
where the second part of the equation is valid only on--shell.
Therefore,
to order ${\lambda}^{-1}$, the conservation law for the $zz$ and $z
\bar{z}$
components of the stress energy--tensor of the theory is obtained,
\be
\bar{\partial} \left( {\partial u \partial \bar{u} \over 1 -
{\mid u \mid}^{2}} \right) + g ~ \partial {\mid u \mid}^{2} = 0.
\ee
At the conformal point $g=0$, this reduces to the chiral conservation
law
for the $zz$ component of the stress--energy tensor of the
$SU(2)/U(1)$
coset model. In the static limit, the result $\partial H = 0$
is also recovered.

Iteration of the recursive relations for $j \geq 2$ yields
higher order non--chiral conservation laws, all depending on the
coupling
constant $g$. In the static limit they all reduce to the conservation
of
$H$ and $Q$, but in the field theory case we are considering now they
turn out to be functionally independent and hence new. To understand
their
nature, in general, it is convenient to
compare them with the infinitely many
symmetries of the $SU(2)/U(1)$ coset model with $g=0$. In the latter
case,
it is known that there is a chiral $W_{\infty}$ symmetry, whose
generators
are bilinear in the parafermions of the model [29--31]. In
particular,
introducing the parafermion currents
\be
\psi_{+} = {\partial u \over \sqrt{1 - {\mid u \mid}^{2}}} ~ V_{+} ~
,
{}~~~ \psi_{-} = {\partial \bar{u}  \over \sqrt{1 - {\mid u
\mid}^{2}}} ~
V_{-},
\ee
where
\be
V_{\pm} = exp \left( \pm {1 \over 2} \int dz ~ J - d \bar{z} ~
\bar{J} \right)
\ee
are defined in terms of the non--chiral $U(1)$ current of the theory,
the
$W_{\infty}$ generators (in a quasi--primary basis and up to an
overall
normalization) are
\be
W_{s} = \sum_{k=0}^{s-2} {{(-1)}^{k} \over s-1} {s-1 \choose k+1}
{s-1 \choose s-k-1} {\partial}^{k} \psi_{+} {\partial}^{s-k-2}
\psi_{-},
\ee
with $ s= 2, ~ 3, ~ 4, ~ \cdots ~ $. For $g=0$, $\psi_{\pm}$ and
$W_{s}$
are all chirally conserved, provided  that the classical
equations of motion of the $SU(2)/U(1)$ coset model are satisfied.
For $g \neq 0$ this is not true, but it still makes
sense to define parafermion currents $\psi_{\pm}$ as in (4.35) and
(4.36),
because the potential ${\mid u \mid}^{2}$ is $U(1)$--invariant.

The conservation laws that result from the
diagonalization of the gauge connections of the perturbed
$SU(2)/U(1)$ coset model can be written systematically for all $g$,
using the
off-critical generalization of the $W_{\infty}$ generators (4.37) in
terms
of the parafermions $\psi_{\pm}$. Note that for
$g \neq 0$, $\bar{\partial} W_{s}$ can not be always brought in the
form
$\partial X$ for appropriately chosen $X$. Although this is
true for $s = 2$, for higher spin currents it is not so.
To describe the conserved currents of the theory in terms of
$\{W_{s}\}$ off--criticality, we have to introduce
appropriate polynomial combinations
of the generators, depending on $s$.
It is the abelianization method that provides the algorithm for
writing
down these higher (non--chiral) conservation laws of the theory in
the
form (4.17).

We find that for $j \geq 2$, the trace of $\tilde{A}_{j}$ is
a total derivative of currents, which are composed from the
subleading componets $h_{j-1}^{(1)} - h_{j-1}^{(2)}$, $h_{j-2}^{(1)}
-
h_{j-2}^{(2)}$, etc. Hence, the only functionally independent
conservation
laws are obtained by considering
$h_{j}^{(1)} - h_{j}^{(2)}$. For example,
for the first few values of $j$ (apart from the ones already
discussed)
we find
\be
h_{2}^{(1)} + h_{2}^{(2)} = - {1 \over 4} ~ \partial (h_{1}^{(1)} -
h_{1}^{(2)}),
\ee
\be
h_{3}^{(1)} + h_{3}^{(2)} = - {1 \over 2} ~ \partial (h_{2}^{(1)} -
h_{2}^{(2)}),
\ee
\be
h_{4}^{(1)} + h_{4}^{(2)} = - {1 \over 32} ~ \partial \left( 24
(h_{3}^{(1)}
- h_{3}^{(2)}) - {(h_{1}^{(1)} - h_{1}^{(2)})}^{2} - {\partial}^{2}
(h_{1}^{(1)} - h_{1}^{(2)}) \right)
\ee
and so on.
As for the $h_{j}^{(1)} - h_{j}^{(2)}$ components
of $\tilde{A}_{j}$, the results of the calculation
are summarized in the appendix for the first few values of $j$.
The complexity of the expressions increases
considerably for higher values of $j$, but they are all calculable
order
by order. Unfortunately, no closed expression for arbitrary $j$ is
available at the moment.
The form of $h_{j}^{(1)} -
h_{j}^{(2)}$ indeed simplifies considerably, when they are written
in terms of the parafermions off-criticality. Extending the
definition of
$W_{s}$ to all values of the coupling constant $g$ and introducing
the
explicit dependence of $A_{0}$ on $u$ and $\bar{u}$,
we obtain after some lengthy
calculation
\be
h_{1}^{(1)} - h_{1}^{(2)} = W_{2},
\ee
\be
h_{2}^{(1)} - h_{2}^{(2)} = - {1 \over 4} ~ W_{3},
\ee
\be
h_{3}^{(1)} - h_{3}^{(2)} = {1 \over 20} ~ \left( W_{4} + 5
{(W_{2})}^{2}
+ {3 \over 2} ~ {\partial}^{2} W_{2} \right),
\ee
\be
h_{4}^{(1)} - h_{4}^{(2)} = -{1 \over 112} ~ (W_{5} + 21 W_{2} W_{3}
+
6 ~ {\partial}^{2} W_{3} )
\ee
and so on. This summary shows that $h_{j}^{(1)} -
h_{j}^{(2)}$ are functionally related to the $W_{s}$ currents, as
advertized. Note that these polynomial combinations are independent
of
$g$.

The form of the infinitely many conservation laws of
the theory (with arbitrary $g$)
is determined completely, by performing the gauge transformation $T$
on the
gauge connection $B$ as well. The iteration of the recursive
relations
also determines $\{t_{i}\}$ order by order. Explicit results
for the first few values of $j$ are given in the appendix, together
with
$\tilde{B}_{j}$. It can be verified that the matrices
$\tilde{B}_{j}$ are all diagonal on--shell, as required by the
abelianization procedure. The final result for the local conservation
laws of the theory (discarding total derivative
terms and overall normalization factors) is
\be
\bar{\partial} W_{3} + g ~ \partial (u \partial \bar{u} - \bar{u}
\partial u) = 0,
\ee
\be
\bar{\partial} (W_{4} + 5 {W_{2}}^{2}) + g ~ \left( {\partial}^{2}
{\mid u \mid}^{2} + 5 ~ {2 {\mid u \mid}^{2} - 1 \over 1 - {\mid u
\mid}^{2}} \partial u \partial \bar{u} \right) = 0,
\ee
\ba
\bar{\partial} (W_{5} + 21 W_{2} W_{3}) & + & g ~ \partial \left(
u {\partial}^{3} \bar{u} - \bar{u} {\partial}^{3} u -
{6 - 13 {\mid u \mid}^{2} \over 1 - {\mid u \mid}^{2}} (\partial u
{\partial}^{2} \bar{u} - \partial \bar{u} {\partial}^{2} u) \right.
\nonumber \\
   & + & \left. 7 ~ {2 - 3 {\mid u \mid}^{2} \over {(1 - {\mid u
\mid}^{2}
)}^{2}} (u \partial \bar{u} - \bar{u} \partial u) \partial u
\partial \bar{u} \right) = 0.
\ea
The higher order conservations laws can be constructed in a similar
fashion, but it is computationally difficult to find closed
expressions
for them, in general. Of course, we also have a complementary set
of non--chirally conserved currents which is obtained by
interchanging
$u \leftrightarrow \bar{u}$ and $\partial \leftrightarrow
\bar{\partial}$.

In the conformal limit $g \rightarrow 0$, the chiral $W_{\infty}$
algebra
is recovered, but in a polynomial basis. Its commutation relations
are
not linear in this basis, but we know that there exists the
quasi--primary basis (4.37) where the linear structure of the algebra
is
manifest. For $g \neq 0$ it does not make sense to talk about the
algebra
of the non--chiral currents, because the symmetries they generate are
global; it is only at the conformal point that global symmetries are
promoted
to local. What makes sense, however, is
to consider the corresponding charges for
all $s = 1, ~ 2, ~ 3, ~ \cdots ~ $, which are conserved and in
involution.
Since they are infinitely many of them, we have a rigorous method
(and an
algorithm) for establishing the complete integrability of the
parafermion
model perturbed by its first thermal operator, in the large $N$
limit.

The path ordered exponential
\be
P ~ exp \left( \int_{P_{1}}^{P_{2}} dz A + d \bar{z} B \right)
\ee
is independent of the path joining two space--time points $P_{1}$ and
$P_{2}$, provided that $A$, $B$ satisfy the zero curvature condition
(4.7). Since the expression (4.48) depends
only on the end points $P_{1}$ and $P_{2}$, we may use it to obtain
an
alternative description of the conserved charges, by considering
paths
joining two spatially far apart points at different times. The
transformation to the diagonal gauge connections $\tilde{A}$,
$\tilde{B}$
makes the path ordering uneccessary. Then, the $1/\lambda$ expansion
of
the new variables yields the infinite set of conservation laws we
have
already discussed. The generating function for the
conservation laws can be obtained from the transition matrix of the
theory,
using the inverse scattering method (see [32] and references
therein).
This method has been applied to the model (1.4) by Lund [13] and
Kulish [33], in order to construct the corresponding action--angle
variables. In this regard, our results are complementary to theirs,
but more
explicit. The main point of this section was the realization that
$W_{\infty}$ and the off--critical generalization of its generators
in the
parafermionic representation (4.37) determine the structure of the
higher
spin conservation laws completely.

The systematic construction of the conserved currents was based on
the
Langrangian description of the model and the zero curvature
formulation
of its classical equations of motion, rather than the null--vector
conditions on the parafermion first thermal operator.
The quantum mechanical generalization of these results should be
straightforward, but computationally difficult. The quantum inverse
scattering method could be used to understand the relation with the
null--vector conditions in a systematic way.
We think that the cohomological framework
of Feigin and Frenkel [34] might be also appropriate
for investigating further the quantum mechanical formulation of the
abelianization method.

\section{\bf KP--like Structure of the Currents}
\setcounter{equation}{0}
\noindent
The conservation laws of the perturbed parafermion theory can be
described systematically in terms of the KP hierarchy. We claim that
the currents $h_{j}^{(1)} - h_{j}^{(2)}$, with $j \geq 1$, can be
interpreted as Hamiltonian densities of the KP hierarchy, thus
providing a more direct way for their explicit construction. This
generalizes the well known relation between the conserved densities
of the sine--Gordon model and the $SL(2)$ KdV hierarchy, to the full
coset theory.

Recall that the KP hierarchy is formulated in terms of the
pseudo--differential operator
\be
L = \partial + q_{1} {\partial}^{-1} + q_{2} {\partial}^{-2} +
q_{3} {\partial}^{-3} + \cdots ~ .
\ee
The KP flows are defined to be
\be
\partial_{t_{r}} L = [ {(L^{r})}_{+} ~ , ~ L ] ,
\ee
with $r = 1, ~ 2, ~ 3, ~ \cdots$ (see for instance [35]). These
equations
admit a Hamiltonian description with
\be
{\cal H}_{r} = {1 \over r} ~ res L^{r}
\ee
being the Hamiltonian density of the $r$--th flow. We have explicitly
\be
{\cal H}_{1} = q_{1},
\ee
\be
{\cal H}_{2} = q_{2} + {1 \over 2} ~ \partial q_{1},
\ee
\be
{\cal H}_{3} = q_{3} + {q_{1}}^{2} + \partial (q_{2} + {1 \over 4} ~
\partial q_{1}),
\ee
\be
{\cal H}_{4} = q_{4} + 3 ~ q_{1} q_{2} + {3 \over 4} ~ \partial
(2 q_{3} + {q_{1}}^{2} + \partial q_{2} + {1 \over 6} ~
{\partial}^{2}
q_{1})
\ee
and so on. Moreover, when the flows (5.2) are written in terms of the
composite fields ${\cal H}$, they become
\be
{\partial}_{t_{r}} {\cal H}_{j} = \partial {\Theta}_{r,j} ~ ,
\ee
where ${\Theta}_{r,j}$ are composite fields of $\{q_{i}\}$ depending
on
$r$, $j$. It is for this reason that the charges
$H_{j} = \int {\cal H}_{j}$ with
$j \geq 1$ are all in involution and the KP system is integrable.

The formulation (5.8) of the KP flows should be compared with the
local
conservation laws of the perturbed parafermion theory. In the
parafermion
case, $h_{j}^{(1)} - h_{j}^{(2)}$ plays the role of ${\cal H}_{j}$
and
$\bar{\partial}$ replaces ${\partial}_{t_{r}}$. Therefore, it is
natural
to identify $h_{j}^{(1)} - h_{j}^{(2)}$ with the Hamiltonian
densities
of the KP hierarchy. Of course, the equivalence between the two
theories
is not exact, because in the KP case the $\Theta$'s can be
rewritten as local functionals of the ${\cal H}$'s, while in the
parafermion
model this is not so. To describe the KP--like structure of the
parafermion currents, we identify the generators $W_{s}$ as
\be
W_{s} = q_{s-1} ~ , ~~~~ s = 2, ~ 3, ~ 4, ~ \cdots
\ee
and postulate
\be
h_{j}^{(1)} - h_{j}^{(2)} = {1 \over 2^{j-1}} {\cal H}_{j}.
\ee
This correspondence is understood modulo total derivative terms,
which
are irrelevant anyway in both the KP flows and the parafermion
conservation
laws, since they lead to trivial field redefinitions.

To illustrate the validity of eq.(5.10), we have to use the right
normalization for the generators $W_{s}$. Following earlier work on
the
$W_{\infty}$ symmetry [29--31], we scale $W_{s}$,
as given by eq.(4.37), by multiplication
with
\be
B(s) = q^{s-2} {2^{s-3} s! \over (2s-3)!!}
\ee
and choose $q = - 1/4$. Then, in terms of the rescaled variables,
eqs.(4.41)--(4.44) become
\be
h_{1}^{(1)} - h_{1}^{(2)} = W_{2},
\ee
\be
h_{2}^{(1)} - h_{2}^{(2)} = {1 \over 2} ~ W_{3},
\ee
\be
h_{3}^{(1)} - h_{3}^{(2)} = {1 \over 4} ~ (W_{4} + {W_{2}}^{2} +
{3 \over 10} ~ {\partial}^{2} W_{3}),
\ee
\be
h_{4}^{(1)} - h_{4}^{(2)} = {1 \over 8} ~ (W_{5} + 3 ~ W_{2} W_{3} +
{6 \over 7} ~ {\partial}^{2} W_{3})
\ee
and so on. Comparison with the expressions (5.4)--(5.7) shows that up
to
total derivative terms (which are gauge dependent in the
abelianization
method, anyway), eq.(5.10) is correct provided that $W_{s} = q_{s-1}$
after
the rescaling. We have verified eq.(5.10) for higher values of $j$ as
well.

The relation we find between the currents of the perturbed
parafermion
theory and the Hamiltonian densities of the KP hierarchy, generalizes
the results we already know for the conservation laws of the
sine--Gordon model. The sine--Gordon model arises as a special case
of the perturbed parafermion theory, obtained for $u = \bar{u}$.
As we have already mentioned, this model describes only the neutral
sector of
our general theory, since both components $J$, $\bar{J}$ of the
$U(1)$
current are identically zero for $u = \bar{u}$. All higher spin
currents
$W_{s}$ with odd values of $s$ also vanish
in this sector, as it can be readily
seen from eq.(4.37). The only non--trivial conservation laws have
even spin. Setting $u = \bar{u} = cos \theta$, we obtain
${\psi}_{+} = {\psi}_{-} = - \partial \theta$. The conservation law
(4.34) for $j = 2$ reduces to
\be
\bar{\partial} \left( {(\partial \theta)}^{2} \right) + g ~
\partial ({cos}^{2} \theta) = 0.
\ee
For $j = 4$, eq.(4.46) becomes
\be
\bar{\partial} \left(5{(\partial \theta)}^{4} - 3{({\partial}^{2}
\theta)}^{2} + 2 ~ \partial \theta {\partial}^{3} \theta \right) + g
{}~
\partial \left(3~cos2 \theta {(\partial \theta)}^{2} - sin 2 \theta ~
{\partial}^{2} \theta \right) = 0.
\ee
Similar expressions result for (even) higher values of $j$, which are
all consistent with the sine--Gordon equation (1.7).

It is well known that the currents $h_{j}^{(1)} - h_{j}^{(2)}$ of the
sine--Gordon model, coincide with the Hamiltonian densities of the
KdV
hierarchy (see for instance [8]). In this case, the
pseudo--differential operator $L$ in eq.(5.1) satisfies the condition
$L^{2} = {\partial}^{2} + v$ and $q_{i}$ are all
functionally dependent on the KdV
field $v$. We have explicitly $q_{1} = v/2$, $q_{2} =
- \partial v /4$, $q_{3} = ({\partial}^{2} v - v^{2})/8$, etc. The
relation
between $v$ and $\theta$ is given by the Miura map. Flows with odd
values of $j$ are trivial, since ${\cal H}_{j}$ become total
$\partial$--derivatives. For the even values of $j$, the KdV
reduction of
eq.(5.10) describes in fact the relation between the conservation
laws
of the two integrable theories.

The validity of eq.(5.10), not only in the neutral sector $u =
\bar{u}$,
but in the full perturbed parafermion theory with arbitrary $U(1)$
charge, suggests that there should be a 2--component reduction of the
KP hierarchy, which actually describes the structure of the currents
$h_{j}^{(1)} - h_{j}^{(2)}$ completely. The 2--component reduction in
question is the non--linear Schrodinger hierarchy, as it has been
pointed out recently [36], using a somewhat different method. The
Lagrangian description of the parafermion coset perturbed by the
first thermal operator is the essential ingredient that links our
work
with theirs. The non--linear Schrodinger hierarchy is practically the
same as the 2--boson realization of the KP hierarchy [37]. The latter
was defined following earlier work on the bosonic realization of
$W_{\infty}$--type algebras [38].

We also point out for completeness that a close relation between the
non--linear Schrodinger equation and vortex dynamics was
discovered several years ago [39], along different lines; vortex
filaments were intepreted as solitons. This result should not be
surprizing, given the fact that the perturbed parafermion coset (with
$g > 0$) in 2--dim Minkowski space describes the relativistic motion
of vortices in constant external field, via the Lund--Regge
formalism.
We think, however, that the role of the non--linear Schrodinger
equation in vortex dynamics should be investigated further and in
connection with perturbed conformal field theories.

We note finally, as a side remark, that it would be interesting to
examine
the Hamiltonian structure of the local conservation laws
of the perturbed coset model.
In analogy with the KP hierarchy, there might exist a Poisson
bracket so that $\bar{\partial} \tilde{A}_{j} + \partial
\tilde{B}_{j} =
0$ could be written as
\be
\bar{\partial} \tilde{A}_{j} = \{\Omega_{j} ~ , \tilde{A}_{j} \},
\ee
when $j \geq 1$ and $g \neq 0$, for appropriately chosen Hamiltonian
$\Omega_{j}$. $\Omega_{j}$ might have a natural interpretation in the
$SU(2)/U(1)$ coset model, which could clarify the structure and
physical
interpretation of the higher spin conservation laws even further.

\section{\bf Discussion and Conclusions}
\setcounter{equation}{0}
\noindent
In this paper we have studied systematically the Lagrangian
description,
physical interpretation and geometry of the parafermion coset model
perturbed by its first thermal operator. The present work is entirely
confined at the classical level and in the large $N$ limit of the
theory. Quantum mechanical aspects and $1/N$ corrections are
interesting
to consider and we hope to return to them in a separate publication.
The main tool in our study has been provided by the geometric
framework of Lund and Regge, upon analytic continuation of the coset
theory in 2--dim Minkowski (base) space. Reviving these
geometric ideas in present day conformal field theories seems to be
advantageous for understanding the Lagrangian description of
their integrable perturbations. Generalizations of these results to
other coset models are certainly of great interest. It seems that the
main problem in the general case is
to understand the zero curvature formulation
of the gauged WZW model (with or without perturbations) and its
geometric interpretation as Gauss--Codazzi integrability conditions
for an embedding problem.

Many of the issues considered here have been studied before, but with
different motivations in mind. Their relevance in 2--dim field theory
and string theory might be quite general, not limited to the present
models. We expect that the correspondence between different string
backgrounds and integrable 2--dim field theories could be developed
further. Then, a systematic approach to the problem of hidden
symmetries
in string theory might result, as advocated in ref.[15].
We conclude with some
applications of the results we have obtained so far.

The infinitely many conservation laws of the perturbed parafermion
model
have a natural geometric interpretation in terms of embeddings. To
treat
all cases together, irrespectively of the sign of the coupling
constant
$g$, we summarize the expressions (3.12) and (3.27) for the extrinsic
curvature as follows
\be
K_{+} \equiv 2~(K_{\s \s} + K_{\s \tau}) = i~ {u \partial \bar{u} -
\bar{u} \partial u \over \mid u \mid \sqrt{1-{\mid u \mid}^{2}}} -
2 \kappa c \mid u \mid \sqrt{1 - {\mid u \mid}^{2}},
\ee
\be
K_{-} \equiv 2~(K_{\s \s} - K_{\s \tau}) = i~ {u \bar{\partial}
\bar{u} -
\bar{u} \bar{\partial} u \over \mid u \mid \sqrt{1 - {\mid u
\mid}^{2}}} -
2 \kappa c \mid u \mid \sqrt{1 - {\mid u \mid}^{2}}.
\ee
The parameter $\kappa$ is a step function: for $g < 0$ it is $\kappa
= 0$,
while for $g \geq 0$ we have $\kappa = 1$ and $c^{2} = g$. The metric
(3.1)
of the 2--dim Euclidean surface $S$ that arises in the geometric
description
of the theory via embeddings, has components
\be
g_{\s \s} = {\mid u \mid}^{2} ~ , ~~~~ g_{\tau \tau} =
1 - {\mid u \mid}^{2}.
\ee
It is also convenient to introduce the notation $ g_{-} = g_{\s \s} -
g_{\tau \tau}$.

It is easy to verify that the conservation law of the $U(1)$
current (4.29) reads as
\be
\bar{\partial} \left( \sqrt{{g_{\s \s} \over g_{\tau \tau}}}~ K_{+} +
\kappa
c ~ g_{-} \right) + \partial \left( \sqrt{{g_{\s \s} \over g_{\tau
\tau}}}
{}~ K_{-} + \kappa c ~ g_{-} \right) = 0.
\ee
The conservation law of the energy--momentum tensor becomes
\be
\bar{\partial} \left( {\left( K_{+} + 2 \kappa c \sqrt{det g}
\right)}^{2}
+ {1 \over 4} ~
{{(\partial g_{-})}^{2} \over det g} \right) +
2 g ~ \partial g_{-} = 0,
\ee
while similar (but considerably more complicated) expressions can be
obtained
for the higher spin currents by rewriting the generators $W_{s}$ in
terms of
the extrinsic curvature and the metric. The calculation is
straightforward
and yields step by step the infinitely many integrals of the
underlying embedding problem. Their dependence on the original field
variables is obtained using the defining relations for the metric and
the
extrinsic curvature tensors in terms of $\eta_{i} (\s , \tau)$ or
$X_{i} (\s , \tau)$, depending on the sign of
the coupling constant $g$. This geometric
interpretation is advantageous for understanding the physical meaning
of
the infinitely many charges associated with the classical evolution
of the
reduced $O(4)$ non--linear $\s$--model or the string propagation in a
4--dim axionic background of vortex type respectively.

For $g = 0$, the currents $W_{s}$ become chiral and generate the
$W_{\infty}$ algebra. This acts as a hidden symmetry on the
transverse
modes of the 4--dim free bosonic string in the orthonormal gauge.
More
details on $W_{\infty}$ Backlund transformations in string theory can
be found in ref.[15]. The lesson for string theory (with or without
axionic background) is that the classical description
of its transverse modes as a
perturbed $SU(2)/U(1)$ coset model (with $g > 0$ or $g = 0$
respectively), implies
an infinite number of conservation laws at the classical level. It
would be
interesting to study the quantum mechanical implications of this
result
and try to generalize the correspondence between
integrable systems living in the string world--sheet and string
dynamics,
to more arbitrary backgrounds in four and higher dimensions.
Also, the relevance of this formalism to the
quantum theory of vortices in (relativistic) superfluids should be
investigated further, using the perturbed gauged WZW model.
Generalization
to other integrable perturbations and higher rank coset models should
be
interesting as well.

Finally, it is straightforward to apply our results to the reduced
$U(2)$
Gross--Neveu model. As was explained in the introduction, in the
special
frame (1.8), (1.9), the local conservation laws of this model can be
obtained from the perturbed parafermion coset by the simple
substitution
(1.10). Generalization to the $U(N)$ case is an interesting problem
which
might be related to integrable perturbations of Grassmannian coset
models,
$SU(2)/U(1)$ being the simplest one.
The $1/N$ expansion of the Gross--Neveu model is
particularly interesting in this framework, since $\bar{\psi}^{\a}
\psi^{\a}$ develops a non--vanishing
vacuum expectation value which breaks chiral
invariance. This result might have a useful
interpretation in quantum coset model calculations.

\vskip2cm
\centerline{\bf Acknowledgments}
\vskip .5cm
\noindent
I am grateful to E. Kiritsis for many useful discussions and
collaboration
on some of the issues considered here, to E. Floratos for
encouragement
and T. Tomaras for constructive critisism, all very much needed after
my 17
month long military service. I also thank A. Polychronakos for a
critical
reading of the manuscript.

\newpage
\centerline{\bf APPENDIX}
\vskip .2cm
\setcounter{section}{0}
\setcounter{equation}{0}

\renewcommand{\theequation}{A.\arabic{equation}}

\noindent
In this appendix
we summarize the results of the iteration of the recursive relations
(4.25)--(4.27) for the first few $h_{j}^{(1)} - h_{j}^{(2)}$,
in terms of the matrix elements of $A_{0}$.
\be
h_{2}^{(1)} - h_{2}^{(2)} = A_{0}^{11} A_{0}^{12} A_{0}^{21} + {1
\over 4} ~
(A_{0}^{12} ~ \partial A_{0}^{21} - A_{0}^{21} ~ \partial
A_{0}^{12}),
\ee
\ba
h_{3}^{(1)} - h_{3}^{(2)} & = & {1 \over 4} ~ A_{0}^{12} A_{0}^{21}
\left( A_{0}^{12} A_{0}^{21} - 4 {(A_{0}^{11})}^{2} \right)
- {1 \over 8} ~ (A_{0}^{12} ~ {\partial}^{2} A_{0}^{21} + A_{0}^{21}
 ~ {\partial}^{2} A_{0}^{12}) \nonumber \\
  & - & {1 \over 2} ~ A_{0}^{11} (A_{0}^{12} ~ \partial A_{0}^{21} -
A_{0}^{21} ~ \partial A_{0}^{12}),
\ea
\ba
h_{4}^{(1)} & - & h_{4}^{(2)}  =  A_{0}^{11} A_{0}^{12} A_{0}^{21}
\left( {(A_{0}^{11})}^{2} - {3 \over 4} ~ A_{0}^{12} A_{0}^{21}
\right) +
{3 \over 16} ~ A_{0}^{12} A_{0}^{21} (A_{0}^{21} ~ \partial
A_{0}^{12} -
A_{0}^{12} ~ \partial A_{0}^{21}) \nonumber \\
  & + & {3 \over 4} ~ A_{0}^{11} \left( A_{0}^{12} ~ \partial
(A_{0}^{11}
A_{0}^{21}) - A_{0}^{21} ~ \partial (A_{0}^{11} A_{0}^{12}) \right) +
{1 \over 16} ~ ( A_{0}^{12} ~ {\partial}^{3} A_{0}^{21}
- A_{0}^{21} ~ {\partial}^{3} A_{0}^{12}) \nonumber \\
& + & {1 \over 8} \left( 2 A_{0}^{12} A_{0}^{21} ~ {\partial}^{2}
A_{0}^{11} + 3 (\partial A_{0}^{11}) \partial (A_{0}^{12} A_{0}^{21})
+ 3 A_{0}^{11} (A_{0}^{12} ~
{\partial}^{2} A_{0}^{21} + A_{0}^{21}
{}~ {\partial}^{2} A_{0}^{12}) \right)
\ea

\noindent
Similarly for $t_{i}$ we have
\be
t_{2}^{12} = {1 \over 2} ~ A_{0}^{11} A_{0}^{12} - {1 \over 4} ~
\partial A_{0}^{12},
\ee
\be
t_{2}^{21} = -{1 \over 2} ~ A_{0}^{11} A_{0}^{21} - {1 \over 4} ~
\partial A_{0}^{21},
\ee
\ba
t_{3}^{12} & = & - {1 \over 2} ~ A_{0}^{12} \left( {(A_{0}^{11})}^{2}
-
{1 \over 4} A_{0}^{12} A_{0}^{21} \right) + {1 \over 4} A_{0}^{12}
{}~ \partial A_{0}^{11} + {1 \over 2} A_{0}^{11} ~ \partial
A_{0}^{12} -
{1 \over 8} {\partial}^{2} A_{0}^{12}, \\
t_{3}^{21} & = & {1 \over 2} A_{0}^{21}
\left( {(A_{0}^{11})}^{2} - {1 \over 4}
A_{0}^{12} A_{0}^{21} \right) + {1 \over 4} A_{0}^{21}~\partial
A_{0}^{11}
+ {1 \over 2} A_{0}^{11} ~ \partial A_{0}^{21} + {1 \over 8}
{\partial}^{2} A_{0}^{21},
\ea
\ba
t_{4}^{12} & = & {1 \over 2} ~ A_{0}^{11} A_{0}^{12}
\left( {(A_{0}^{11})}^{2}
- {3 \over 4} A_{0}^{12} A_{0}^{21} \right) - {3 \over 4} ~
A_{0}^{11}
{}~ \partial (A_{0}^{11} A_{0}^{12}) \nonumber \\
    & + & {1 \over 16} ~ A_{0}^{12} (A_{0}^{12} ~ \partial A_{0}^{21}
+ 4
A_{0}^{21} ~ \partial A_{0}^{12}) -
{}~ {1 \over 16} {\partial}^{3} A_{0}^{12}
\nonumber \\
    & + & {1 \over 8} \left(3 A_{0}^{11} ~ {\partial}^{2} A_{0}^{12}
+
3 (\partial A_{0}^{11})(\partial A_{0}^{12}) + A_{0}^{12} ~
{\partial}^{2}
A_{0}^{11} \right),
\ea
\ba
t_{4}^{21} & = & - {1 \over 2} ~ A_{0}^{11} A_{0}^{21} \left(
{(A_{0}^{11})}^{2} - {3 \over 4}
A_{0}^{12} A_{0}^{21} \right) - {3 \over 4} ~
A_{0}^{11} ~ \partial (A_{0}^{11} A_{0}^{21}) \nonumber \\
   & + & {1 \over 16} ~ A_{0}^{21} (A_{0}^{21} ~ \partial A_{0}^{12}
+
4 A_{0}^{12} ~ \partial A_{0}^{21}) - {1 \over 16} ~ {\partial}^{3}
A_{0}^{21} \nonumber \\
   & - & {1 \over 8} \left( A_{0}^{21} ~ {\partial}^{2} A_{0}^{11} +
3(\partial A_{0}^{11})(\partial A_{0}^{21}) + 3 A_{0}^{11}
{}~ {\partial}^{2} A_{0}^{21} \right).
\ea

\noindent
For $\tilde{B}_{j}$ we have
\be
\tilde{B}_{2} = [B_{1}, t_{1}] + [B_{0}, t_{2}] + t_{1} [t_{1},
B_{0}]
- \bar{\partial} t_{2} + t_{1} \bar{\partial} t_{1},
\ee
\ba
\tilde{B}_{3} & = & [B_{0}, t_{3}] + [B_{1}, t_{2}] + t_{1}
\left( [t_{2}, B_{0}]
+ [t_{1} , B_{1}] \right) \nonumber \\
   & - & \bar{\partial} t_{3} + t_{1} \bar{\partial}
t_{2} + \left( t_{2} - {(t_{1})}^{2} \right) (\bar{\partial} t_{1}
+ [t_{1} , B_{0}]),
\ea
\ba
\tilde{B}_{4} & = & [B_{0}, t_{4}] + [B_{1}, t_{3}] + t_{1} [t_{3},
B_{0}]
+ t_{1} [t_{2}, B_{1}] + t_{1} \bar{\partial} t_{3}
\nonumber \\
  & + & \left( t_{2} - {(t_{1})}^{2} \right) \left( [t_{2}, B_{0}] +
[t_{1}, B_{1}] + \bar{\partial} t_{2} \right) - \bar{\partial} t_{4}
\nonumber \\
  & + & \left(t_{3} - t_{1} t_{2} - t_{2} t_{1} +
{(t_{1})}^{3} \right) ([t_{1}, B_{0}] + \bar{\partial} t_{1}).
\ea

\newpage
\centerline{\bf REFERENCES}
\begin{enumerate}
\item A. Zamolodchikov, Sov. Phys. JETP Lett. \underline{46} (1987)
160;
Int. Jour. Mod. Phys. \underline{A3} (1988) 743.
\item A. Zamolodchikov, in {\em ``Integrable Systems in Quantum Field
Theory and Statistical Mechanics"}, Adv. Stud. Pure Math.
\underline{19}
(1989) 1.
\item J. Wess and B. Zumino, Phys. Lett. \underline{B37} (1971) 25;
E. Witten, Comm. Math. Phys. \underline{92} (1984) 455;
D. Karabali, Q--Han Park, H. Schnitzer and Z. Yang, Phys. Lett.
\underline{B216} (1989) 307; D. Karabali and H. Schnitzer, Nucl.
Phys.
\underline{B329} (1990) 649; K. Gawedski and A. Kupiainen, Nucl.
Phys.
\underline{B320} (1989) 625.
\item K. Bardacki, M. Crescimanno and E. Rabinovici, Nucl. Phys.
\underline{B344} (1990) 344; A. Giveon, Mod. Phys. Lett.
\underline{A6}
(1991) 2843; E. Kiritsis, Mod. Phys. Lett. \underline{A6} (1991)
2871.
\item E. Witten, Phys. Rev. \underline{D44} (1991) 314; G. Mandal,
A. Sengupta and S. Wadia, Mod. Phys. Lett. \underline{A6} (1991)
1685;
S. Elitzur, A. Forge and E. Rabinovici, Nucl. Phys. \underline{B359}
(1991) 581.
\item V. Fateev and A. Zamolodchikov, Int. Jour. Mod. Phys.
\underline{A5} (1990) 1025;
V. Fateev, Int. Jour. Mod. Phys. \underline{A6} (1991) 2109;
H. de Vega and V. Fateev, Int. Jour. Mod. Phys. \underline{A6} (1991)
3221.
\item V. Fateev and Al. Zamolodchikov, Phys. Lett. \underline{B271}
(1991) 91.
\item V. Drinfeld and V. Sokolov, Jour. Sov. Math. \underline{30}
(1985)
1975.
\item D. Olive and N. Turok, Nucl. Phys. \underline{B257(FS14)}
(1985)
277, \underline{B265(FS15)} (1986) 469.
\item T. Eguchi and S. Yang, Phys. Lett. \underline{B224} (1989) 373.
\item T. Hollowood and P. Mansfield, Phys. Lett. \underline{B226}
(1989)
73.
\item F. Lund and T. Regge, Phys. Rev. \underline{D14} (1976) 1524;
F. Lund, Phys. Rev. Lett. \underline{38} (1977) 1175.
\item F. Lund, Phys. Rev. \underline{D15} (1977) 1540.
\item K. Pohlmeyer, Comm. Math. Phys. \underline{46} (1976) 207.
\item I. Bakas, ``$W_{\infty}$ {\em Symmetry of the Nambu--Goto
String
in 4 Dimensions"}, preprint CERN--TH.7046, October 1993.
\item Y. Nambu and G. Jona--Lasinio, Phys. Rev. \underline{122}
(1961)
345; D. Gross and A. Neveu, Phys. Rev. \underline{D10} (1974) 3235.
\item A. Neveu and N. Papanicolaou, Comm. Math. Phys. \underline{58}
(1978) 31.
\item S. Coleman, Phys. Rev. \underline{D11} (1975) 2088; S.
Mandelstam,
Phys. Rev. \underline{D11} (1975) 3026.
\item A. Zamolodchikov and V. Fateev, Sov. Phys. JETP \underline{62}
(1985)
215; Sov. Jour. Nucl. Phys. \underline{43} (1986) 657.
\item V. Knizhnik and A. Zamolodchikov, Nucl. Phys. \underline{B247}
(1984) 83.
\item D. Gepner and E. Witten, Nucl. Phys. \underline{B278} (1986)
493.
\item R. Dijkgraff, E. Verlinde and H. Verlinde, Nucl. Phys.
\underline{B371} (1992) 379;
I. Antoniadis, C. Bachas, J. Ellis and D. Nanopoulos, Phys. Lett.
\underline{B211} (1988) 393; A. Tseytlin, Nucl. Phys.
\underline{B399}
(1993) 601;
I. Bars and K. Sfetsos, Phys. Rev. \underline{D46} (1992) 4510.
\item S. Coleman, in {\em ``Aspects of Symmetry: Selected Erice
Lectures"},
Cambridge University Press, Cambridge, 1985.
\item B. Getmanov, Sov. Phys. JETP Lett. \underline{25} (1977) 119;
Theor. Math. Phys. \underline{38} (1979) 124.
\item R. Koberle and J. Swieca, Phys. Lett. \underline{B86} (1979)
209.
\item L. Eisenhart, {\em ``Riemannian Geometry"}, Princeton
University
Press, Princeton, New Jersey, 1964.
\item M. Kalb and P. Ramond, Phys. Rev. \underline{D9} (1974) 2273.
\item M. Rasetti and T. Regge, Physica \underline{80A} (1975) 217.
\item I. Bakas, Phys. Lett. \underline{B228} (1989) 57; Comm. Math.
Phys.
\underline{134} (1990) 487.
\item C. Pope, L. Romans and X. Shen, Phys. Lett.
\underline{B236} (1990) 173; Nucl. Phys. \underline{B339} (1990) 191.
\item I. Bakas and E. Kiritsis, Nucl. Phys. \underline{B343} (1990)
185;
Phys. Lett. \underline{B301} (1993) 49.
\item L. Faddeev, in {\em ``Recent Advances in Field Theory and
Statistical Mechanics"}, Les Houches Lectures, Section {\em XXXIX},
ed. by
J.-B. Zuber and R. Stora, Elsevier Science Publishers B.V., 1984.
\item P. Kulish, Theor. Math. Phys. \underline{33} (1977) 1016.
\item B. Feigin and E. Frenkel, Phys. Lett. \underline{B276} (1992)
79.
\item G. Segal and G. Wilson, Publ. Math. IHES \underline{61} (1985)
5.
\item J. Schiff, {\em ``The Nonlinear Schrodinger Equation and
Conserved Quantities in the Deformed Parafermion and $SL(2, R)/U(1)$
Coset Models"}, preprint IASSNS--HEP--92/57, August 1992; J. Schiff
and
D. Depireux, {\em ``On the Hamiltonian Structures and Reductions of
the KP
Hierarchy"}, preprint IASSNS--HEP--92/66, September 1992.
\item D. Depireux, Mod. Phys. Lett. \underline{A7} (1992) 1825.
\item I. Bakas and E. Kiritsis, Int. Jour. Mod. Phys. \underline{A7}
[Suppl.1A] (1992) 55; F. Yu and Y.-S. Wu, Phys. Rev. Lett.
\underline{68} (1992) 2996.
\item H. Hasimoto, Jour. Fluid Mech. \underline{51} (1972) 477.
\end{enumerate}

\end{document}